\documentclass{aa}

\input psfig.tex

\begin{document}

\newcommand{\ms}{M$_{\odot}$}
\newcommand{\zs}{Z$_{\odot}$}

\title{ Abundance evolution of intermediate mass elements
(C to Zn) in the Milky Way halo and disk}

\author{Aruna Goswami \inst{1} \and Nikos Prantzos \inst{2}}

\institute{Indian Institute of Astrophysics, Bangalore, 560034, India. 
(aruna@iiap.ernet.in)
\and Institut d'Astrophysique de Paris, C.N.R.S., 98 bis Bd.
              Arago,75014 Paris, France. (prantzos@iap.fr)}

\date{Received ; accepted }

\authorrunning{A. Goswami and N. Prantzos}

\titlerunning {Evolution of intermediate mass elements in the 
Milky Way }

\maketitle

\begin{abstract}

We present a comprehensive study of the evolution of the abundances of 
intermediate mass elements, from C to Zn, in the Milky Way halo and in the
local disk. We use a consistent model to describe the evolution of those
two galactic subsystems. The halo and the disk are assumed to evolve 
independently, both starting with gas of primordial composition, and 
in different ways: strong outflow is assumed to take place during the 
$\sim$1 Gyr of the halo formation, while the disk is built by slowly
infalling gas.
This description of the halo+disk evolution can correctly account for the
main observational constraints (at least in the framework of simple models
of galactic chemical evolution). We utilise then metallicity dependant yields
to study the evolution of all elements from C and Zn. Comparing our results
to an extensive body of observational data (including very recent ones), we
are able to make a critical analysis of the successes and shortcomings of
current yields of massive stars. Finally, we discuss qualitatively some 
possible ways to interpret the recent data on oxygen vs iron, which suggest
that oxygen behaves differently from the other alpha-elements.


\end{abstract}


\maketitle

\section{Introduction}
 
In the past ten years or so, progress in our understanding of the chemical 
evolution of the Milky Way came mainly from observations concerning the 
composition of stars in the halo and the local disk. The seminal works of 
Edvardsson et al. (1993) for the disk, and Ryan et al. (1996) and McWilliam 
et al (1995) for the halo (along with many others) provided detailed 
abundance patterns that reveal, in principle, the chemical history of 
our Galaxy.

The interpretation of these data is not straightforward, however, since it
has to be made in the framework of some appropriate model of galactic 
chemical evolution (GCE). 
Only one of the three  main ingredients of GCE models can be calculated
from first principles at present:
the stellar yields. For the other two ingredients, i.e. the stellar initial
mass function (IMF) and the star formation rate (SFR), one has to rely on 
empirical prescriptions.

Considerable progress in GCE studies was made possible after the publication 
of the yields from massive stars of Woosley and Weaver (1995, hereafter 
WW1995). This work made available, for the first time, yields for an
extensive set of isotopes (from H to Zn), stellar masses (from 11 to 40 \ms)
and metallicities (from Z=0 to Z=\zs), making thus possible a detailed
comparison of theory to observations. 
Only two works until now explored fully the potential of the WW1995 yields.
Timmes et al. (1995) adopted a simple GCE model with infall, appropriate for 
the Milky Way disk but certainly not for the halo (see Sec. 3.3); 
in the framework of that model they made a case-by-case assessment of the
strengths and weaknesses of the WW1995 yields, identifying the large
yields of Fe as the main weak point. On the other hand, Samland (1998)
utilised a chemo-dynamical model for the Milky Way evolution
(describing, presumably, correctly the halo and the disk), but introduced
several approximations on the stellar lifetimes and the metallicity dependant
yields of WW1995; he evaluated then the deviation of the published yields
from the ``true'' galactic ones, the latter being derived by a comparison
of his model results with observations of the halo and disk abundance patterns.

Those two works are the only ones that utilised metallicity dependant yields
and studied the full range of intermediate mass chemical elements. Several 
other works focused on specific elements and utilised only metallicity
independant yields (e.g. Pagel and Tautvaisiene 1995; Chiappini et al.
1997, 1999; Thomas et al. 1998 etc.)

In this work we reassess the chemical evolution of the elements from C to Zn in 
the Milky Way, using the WW1995 yields. Our work differs in several aspects
from the one of Timmes et al. (1995) and, in fact, from any other work on that
topic, performed in the framework of simple GCE models: the main novelty is 
that we use appropriate models for {\it both} the halo and the disk, 
correctly reproducing the main observational constraints for those two
galactic subsystems (see Sec. 4). Moreover, we adopt the Kroupa et al. (1993)
IMF, which presumably describes the distribution of stellar mases better 
than the Salpeter IMF (adopted in Timmes et al 1995, Samland 1998, and most
other studies of that kind). Also, w.r.t. the work of Timmes et al. (1995),
our comparison to observations benefits from the wealth of abundance data
made available after the surveys of Ryan et al. (1996), McWilliam 
et al (1995), Chen et al (2000) and many others (listed in Table 1). These
data allow to put even stronger constraints on the stellar yields
as a function of metallicity. We notice that we do not include yields from
intermediate mass stars in our study, since we want to see to what extent
those stars (or other sources) are required to account for the observations.

The plan of the paper is as follows: In Sec. 2 we discuss briefly the uncertainties
currently affecting the yields of massive stars and present the yields of WW1995.
We also present those of a recent work (Limongi et al. 2000), which compare
fairly well to those of WW1995 but show interesting differences for several 
elements. Moreover, we present the recent yields of Iwamoto et al. (1999) for 
supernovae Ia,
calculated for white dwarfs resulting from stars of solar and zero initial 
metallicities, respectively; they are slightly different from the ``classical''
W7 model for SNIa (Thielemann et al. 1986), and we adopt them in our study.
In Sec. 3 we present our chemical evolution model, stressing the 
importance of adopting appropriate ingredients for the halo and the disk.
In Sec. 4 we ``validate'' our model by comparing successfully its results to the
main observational constraints. We also show that current massive star yields 
have difficulties  in explaining the solar composition of Sc, Ti and V.
In Sec. 5 we present the main result of this work, i.e. a detailed comparison
of the model to observations of abundance patterns in halo and disk stars.
This comparison allows to identify clearly  the successes and inadequacies
of the WW1995 yields; some of those inadequacies  may be due to physical 
ingredients not as yet incorporated in ``standard'' stellar models 
(i.e. mass loss or rotationally induced mixing), but the origins of others are 
more difficult to identify. Since the evolution of Fe (usually adopted as
``cosmic clock'') is subject to various theoretical uncertainties  - Fe yields
of massive stars, rate of Fe producing supernovae Ia etc -  we also plot
our results as a function of Ca; 
comparison to available observations (never performed before)
gives then  a fresh and instructive view of the metallicity dependence of 
the massive star yields.
In Sec. 6, we discuss qualitatively some 
possible ways to interpret the recent data of Israelian et al (1998) and
Boesgaard et al (1999) on oxygen vs iron; these data suggest
that oxygen behaves differently from the other alpha-elements and, if
confirmed, will require some important revision of current ideas on
stellar nucleosynthesis. Finally, in Sec. 7 we compare the model evolution
of the Mg isotopic ratios to recent observations of disk and halo stars;
we find that the WW1995 yields underestimate the production of the
neutron-rich Mg isotopes at  low metallicities.

\section {Yields of massive stars and supernova Ia}

Massive stars are the main producers of most of the heavy isotopes in the
Universe (i.e. those with mass number A$>$11). 
Elements up to Ca are
mostly produced in such stars by hydrostatic burning, whereas Fe peak
elements are produced by the final supernova explosion (SNII), as well as
by white dwarfs exploding in binary systems as SNIa. Most of He, C, N and
minor CO isotopes, as well as s-nuclei comes from intermediate mass
stars (2-8 M$_{\odot}$).
A detailed  discussion of the yields of massive stars and their role in
galactic chemical evolution studies has been presented in a recent review
(Prantzos 2000); here we summarize the most important points.

Extensive calculations performed in the 90ies by a few 
groups with 1-D stellar codes (Woosley and Weaver 1995, Arnett 1996, 
Thielemann et al. 1996, Chieffi et al. 1998, Maeder 1992, Woosley et al. 1993,
Aubert et al. 1996, Limongi et al. 2000)
have revealed  several interesting features of nucleosynthesis in massive 
stars.  In particular, 
the structure and composition of the  pre-supernova star reflects the
combined effect of (i) the various mixing mechanisms (convection,
semi-convection, rotational mixing etc.), determining the extent of the
various ``onion-skin'' layers, (ii) the amount of mass-loss 
(affecting mostly the yields of the He and CNO nuclei, present in the
outer layers)  and (iii) the rates of the relevant nuclear ractions
(determining the abundances of the various species in each layer). 

On the other hand, the calculation of the Fe-core collapse supernova explosion
is still one of the major challenges in stellar astrophysics. 
Multi-dimensional hydrodynamical simulations in the 90ies revealed the
crucial role played by neutrino transport in the outcome of the explosion
(e.g. Janka 1998 and references therein). In the absence of a well-defined
explosion scheme, modelers of supernovae nucleosynthesis
have to initiate the explosion somehow (by introducing either an ``internal
energy bomb'', or  a ``piston'', e.g. Aufderheide et al. 1991) 
and adjust the shock energy as to have a pre-determined 
final kinetic energy, usually the ``classical'' value of 10$^{51}$ ergs 
(after accounting for the binding energy of the ejected matter).
This procedure introduces one more degree of uncertainty in
the final yields. Moreover, the ejected amount of   Fe-peak nuclei
depends largely on the position of the {\it mass-cut}, the surface
separating the material falling back onto the neutronized core from the
ejected envelope. The position of this surface depends on the details
of the explosion (i.e. the delay between the bounce and the
neutrino-assisted explosion, during which the proto-neutron star
accretes material) and cannot be evaluated currently with precision
(e.g. Thielemann et al. 1999 and references therein).

\begin{figure*}
\psfig{file=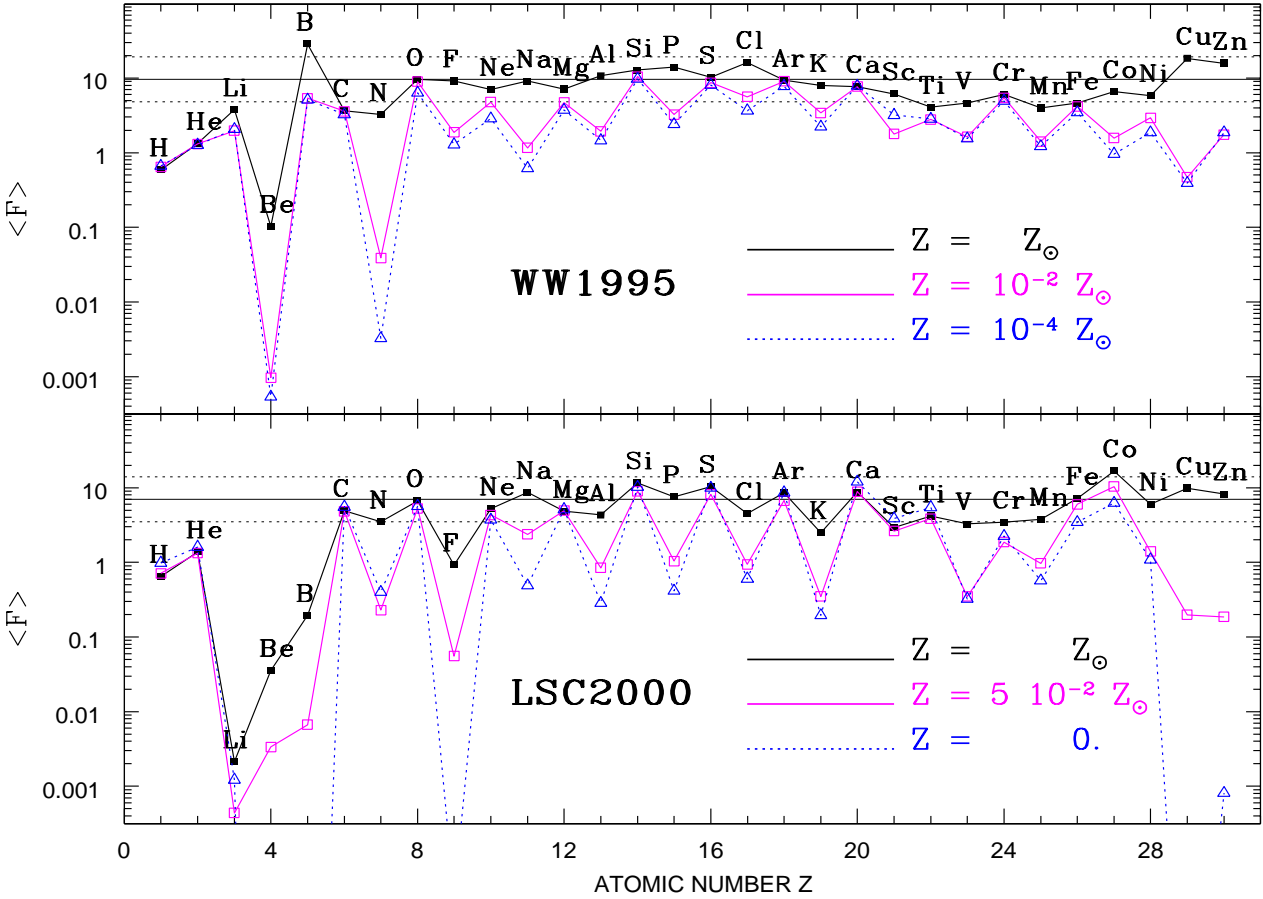,height=13.cm,width=\textwidth,angle=-90}
\caption{\label{} 
Average overproduction factors (over a Kroupa et al. (1993) IMF, 
see Eq. 1) of the yields of
Woosley and Weaver 1995 (WW1995, {\it upper panel}) and Limongi, Straniero 
and Chieffi 2000 (LSC2000, {\it lower panel}) for 3 different initial stellar
metallicities. In both cases, the {\it solid horizontal lines} are 
placed at $F_{oxygen}$ and the two {\it dotted horizontal lines}
at half and twice that value, respectively. The "odd-even effect''
is clearly seen in both the data sets.  N behaves as a pure
"secondary". The elements He, C, N, Li and Be in both cases (as well as
B and F in LSC2000) require another production site.  
}
\end{figure*}

In the light of the aforementioned results,  intermediate mass elements 
produced in massive stars may be divided in three major groups:

(i) C, N, O, Ne, and Mg are mainly produced in hydrostatic burning phases.
They are mostly found in layers which are not heavily processed by 
explosive nucleosynthesis. The yields of these elements depend on the
pre-supernova model (convection criterion, mixing processes, mass loss
and nuclear reaction rates).

(ii) Al, Si, S, Ar and Ca are also produced by hydrostatic burning, but
their abundances are subsequently affected by the passage of the shock 
wave. Their yields depend on both the pre-supernova model and the shock
wave energy.

(iii) Fe-peak elements as well as some isotopes of lighter elements
like Ca, S and Ti are produced by the final SN explosion (SN II).
Their yields depend crucially upon the explosion mechanism and the position 
of the "mass-cut".

The outcome of nucleosynthesis depends also on the   initial metallicity
of the star.   During H-burning the   initial CNO transforms into
$^{14}$N,   which transforms mostly into $^{22}$Ne during He-burning,
through $\alpha$-captures and a $\beta$ decay. The surplus of neutrons
in $^{22}$Ne (10 protons and 12 neutrons) affects the products
of subsequent burning stages, in particular those  of explosive burning.
This neutron surplus increases with initial metallicity and
favours the production of odd nuclei ($^{23}$Na, $^{27}$Al, $^{31}$P etc.),
giving rise to  the so-called  "odd-even" effect.

In the past few years, several groups have reported results of 
pre and post-explosive nucleosynthesis calculations in massive stars
with detailed networks. Thielemann et al. (1996) used bare He cores of 
initial metallicity $Z_\odot$, while Arnett (1996) simulated the
evolution of He  cores  (with polytropic-like trajectories) and
studied different initial metallicities. Full stellar models (neglecting
however, rotation and mass loss ) were studied by Woosley and Weaver (1995,
for masses 12, 13, 15, 18, 20, 22, 25, 30, and 40 $M_\odot$ and
metallicities Z=0, 10$^{-4}$,  10$^{-2}$,  10$^{-1}$, and 1 Z$_\odot$)
and Limongi,  Straniero and Chieffi (2000, for masses 13, 15, 20, 25 $M_\odot$
and metallicities Z=0, 5 10$^{-2}$ and 1 Z$_\odot$). Comparison of the
various yields on a star by star basis shows that there are large 
discrepancies between the different authors (due to differences in the 
adopted physics) although for some elements,
like oxygen, there is a rather good agreement. Moreover,  
the yields do not show a monotonic behaviour with  stellar mass.

Notice that the overall yield used in chemical evolution studies depends on 
both the individual stellar yields and the stellar IMF. Despite a vast amount
of theoretical and observational work, the exact shape of the IMF is 
not  well known yet (Gilmore et al. 1998 and references therein). It is
however clear that the IMF flattens in the low mass range and cannot be 
represented by a power law of a single slope (e.g Kroupa et al. 1993).  
The shape of the IMF  introduces a further uncertainty of a factor 
$\sim$ 2 as to the absolute yield value of each isotope (Wang and Silk 1993).

In Fig. 1 we present the metallicity dependant yields of Woosley and Weaver 
1995 (hereafter WW1995) and Limongi, Straniero and Chieffi 2000 
(hereafter LSC2000), folded with a 
Kroupa et al. (1993) IMF.  
They are presented as {\it overproduction factors}, i.e. 
the yields (ejected mass of a given element) are divided by the 
mass of that element initially present in the part of the star 
that is finally ejected, i.e.
\begin{equation}
<F> \ = \ {{\int_{M1}^{M2} Y_i(M) \ \Phi(M) \ dM} \over {\int_{M1}^{M2} 
X_{\odot,i}(M-M_R) \ \Phi(M) \ dM}}
\end{equation}
where: $\Phi(M)$ is the IMF, $M1$ and $M2$ the lower and upper mass limits
of the stellar models (12 \ms \ and 40 \ms \ for WW1995, 13 \ms \ and 25 \ms
for LSC2000, respectively), $Y_i(M)$ are the individual stellar yields
and $M_R$ the mass of the stellar remnant. Adopting $X_{\odot,i}$ in Eq. (1)
creates a slight inconsistency with the definition of the overpoduction factor
given above, but it allows to visualize the effects of metallicity in the yields
of secondary and odd elements.

From Fig. 1 it can be seen that i) most of the intermediate mass elements are nicely
co-produced (within a factor of 2) in both calculations of solar metallicity
stars; ii) some 
important discrepancies (e.g. Li, B, F) can be readily understood in terms
of neutrino-induced nucleosynthesis, included in the WW1995 but not in the
LSC2000 calculation; iii) the odd-even effect is clearly present in both
calculations, but seems to be more important in LSC2000.
For solar metallicity stars
most of the even Z elements are produced with similar yields in both
calculations, while odd Z elements in LSC2000  are produced with 
systematically lower yields than in WW1995.
A common feature of both calculations is the relative underproduction
of C, N, Sc, V and Ti w.r.t. O. 
C and N clearly require another source (intermediate mass stars and/or Wolf-Rayet
stars, see Prantzos et al. 1994 and Sec. 4.2). The situation is less clear for the
other elements, Sc, V and Ti.

In this work we adopt the metallicity dependant yields of WW1995, keeping
in mind that the use of LSC2000 yields may lead to different results for 
some odd elements. For illustration
purposes we shall also use the WW1995 yields at constant (=solar) metallicity.
There are interesting differences between the two cases (i.e. constant vs. variable
metallicity yields) and this instructive
comparison has never been done before. We notice that in the case of the most 
massive stars (M$>$30 \ms) WW1995 performed 3 calculations, making
different assumptions about the kinetic energy of the supernova ejecta. We adopt
here their set of models A, in which, following 
the explosion, most of the heavy elements
in the inner core fall back to form a black hole of a few solar masses;
because of the form of the IMF, these very massive stars play a negligible role
in shaping the elemental abundance ratios. As stressed
in the Introduction, we consider no yields from intermediate mass stars
in this work; our explicit purpose is to check to what extent
massive stars can account for observations of intermediate mass elements
and for which elements the contribution of intermediate mass stars 
is mandatory.

\begin{figure}
\psfig{file=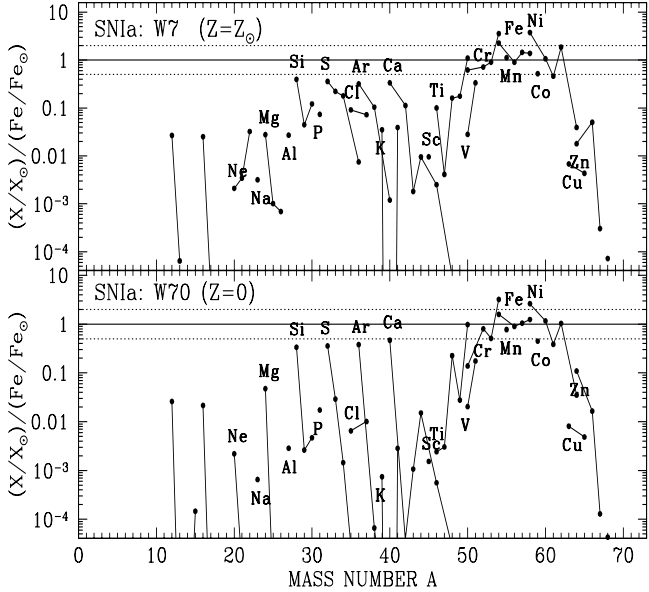,height=8.5cm,width=0.5\textwidth,angle=-90}
\caption{\label{} 
Isotopic yields of SNIa resulting from Chandrasekhar mass white dwarfs,
according to Iwamoto et al. (1999).
{\it Top panel:} model W7  (the white dwarf results from a star of 
initial metallicity Z=Z$_{\odot}$. {\it Bottom panel:} model W70
(the white dwarf results from a star of  initial metallicity Z=0).
Both models are calculated with updated nuclear reaction rates
(with respect to the ``old'' W7 model of Thielemann et al. 1986).
In both cases, the overproduction factor of $^{56}$Fe is taken as 1,
while variations by a factor of 2 are indicated by {\it dotted} lines.
$^{54}$Cr and $^{58}$Ni are clearly overproduced in those models.
 }
\end{figure}

There is a strong observational argument, suggesting that massive stars are not 
the sole producers of Fe peak nuclei in the solar neighbourhood :
the observed decline  in the [O/Fe] ratio (Fig. 3, lower panel) 
from its $\sim$3 times
the solar value in the halo stars ([O/Fe] $\sim$0.5 for [Fe/H]$<$-1) down to 
solar in disk stars. This decline 
is usually interpreted as due to injection  of Fe and Fe group elements
by SN Ia. Assuming that massive stars are the only source of O and Fe in the
halo phase and they produce a ratio of Fe/O$\sim$1/3 solar, the remaining 
$\sim$2/3 of Fe in the late disk should be produced by a late source, 
presumably SNIa. 

The WW1995 yields lead to approximately solar abundance ratios of O/Fe 
(or $\alpha$-element/Fe). This lead Timmes et al. (1995) to suggest that
the Fe yields of WW1995 are probably overestimated. Following their 
suggestion, we adopt here half the nominal values for the WW1995
yields of Fe-peak elements (from Cr to Zn). 
Taking into account the uncertainties
currently affecting those yields, such a reduction is not unreasonable.
Our procedure allows to reproduce the observed O/Fe, but does not alter
the abundance ratios {\it between} Fe-peak elements.
 
To account for the additional source of Fe-peak elements
we utilise the recent yields of SNIa from the exploding white dwarf models 
of Iwamoto et al. (1999). These are updated versions of the original
W7 model of Thielemann et al. (1986).
In this model, the deflagration is starting in the centre of an accreting 
Chandrashekhar-mass CO white dwarf,  burns $\sim$ half of the stellar material in
Nuclear Statistical Equilibrium
and produces $\sim$ 0.7 $M_{\odot}$ of $^{56}$Fe ( in the form of $^{56}$Ni).
It also produces all other Fe-peak isotopes and in particular $^{58}$Ni and $^{54}$Cr. 
This can be seen in Fig. 2, where the overproduction factors (normalised to 
the one of $^{56}$Fe) of the SNIa yields are plotted for two models: one calculated
for a white dwarf resulting from a star with solar initial metallicity (W7) and
another for a white dwarf resulting from a star of zero initial metallicity (W70).
The main difference between the two model results lies in the large 
underproduction
of odd-isotopes in the latter case. In our calculation, we use the yields
of those two models, linearly interpolated as a function of metallicity.

The problem with SNIa is
that, although the current rate of SNIa/SNII is 
constrained by observations in external spiral galaxies
(Tammann et al. 1994), the past
history of that rate (depending on the nature of progenitor 
systems) is virtually unknown. Thus, at present,
it is rather a mystery why the timescale for the onset of SNIa
activity (presumably producing the observed decline of O/Fe in the disk) 
coincides with the timescale for halo formation. An original suggestion was 
recently made in Kobayashi  et al. (1998), whereby  SNIa appear at a rate which is
metallicity dependant;
the interest of this scenario lies in the fact that 
SNIa enter the cosmic scene at just the right moment.
For the purpose of this work, we shall adopt the formalism of Matteucci and Greggio
(1986), adjusting it as to have SNIa appearing mostly after the first Gyr,
i.e. at a time when [Fe/H]$\sim$-1.

At this point we would like to point out that two recent observations
(Israelian et al. 1998 and Boesgaard et al. 1999)  challenged the ``traditional''
view of O vs Fe evolution, by finding a trend
of O/Fe {\it constantly increasing} with decreasing metallicity (open
triangles in Fig. 3). This intriguing trend
is not confirmed by subsequent studies (Fullbright and Kraft 1999), but the question
remains largely open today.
If the new findings are confirmed, some of our ideas on stellar nucleosynthesis 
should be revised. 
Some possibilities of such a revision are explored in Sec. 6.

\section{ The model of galactic chemical evolution }

Models of  chemical evolution for the halo and the disk of the Milky Way 
are constructed adopting the standard  formalism (Tinsley 1980,  Pagel 1997).
The classical set of the equations of galactic chemical
evolution is solved numerically for each zone, without the Instantaneous
Recycling Approximation (IRA). At the star's death its ejecta is assumed to be
thoroughly mixed in the local interstellar medium (instantaneous mixing
approximation), which is then
characterized by a unique composition at a given time. Abundance scatter 
cannot be treated in that framework, and this constitutes an important drawback of
this type of ``classical'' models, since observations suggest a scatter 
of element to element ratios which increases with decreasing metallicity 
(Ryan et al. 1996). The basic ingredients of the model are described below.

\subsection {Stellar lifetimes and remnant masses}

The stellar lifetimes ~$\tau_{M}$ as a function of stellar mass M are taken 
from the work of the Geneva group (Schaller et al. 1992, Charbonnel et al. 1996), 
where the effects of  
mass loss on the duration of H and He burning phases are taken into account.

Stars with mass M$<$9$ M_{\odot}$
are considered to become white dwarfs with mass $M_{R}(M/M_{\odot}) $
=0.1$(M/M_{\odot}$)+0.45 (Iben and Tutukov 1984).
Stars with mass M $>$9$M_{\odot}$
explode as core collapse supernovae leaving behind a neutron star of mass
$M_{R}=1.4 M_{\odot}$ (as suggested by the observations of 
neutron stars in binary systems, e.g.  Thorsett and  Chakrabarty 1999).
The heaviest of those stars may form a black hole, but 
the mass limit for the formation of  stellar black holes
is not known at present and cannot be inferred from theoretical or 
observational arguments (e.g. Prantzos 1994), 
despite occasional claims to the contrary. 
Due to the steeply decreasing stellar Initial
Mass Function in the range of massive stars (see Sec. 3.2), as far as 
the mass limit   for stellar black hole formation is M$_{BH}>$40$M_{\odot}$ 
the results of chemical evolution are not expected to be significantly 
affected by the exact value of M$_{BH}$.

We stress that in our calculations we do take into account the amount
of mass returned in the interstellar medium (ISM) by stars with M$<$11 \ms \
and  M$>$40 \ms \
in the form of H, He, but  also of all heavier elements, up to Zn.
Since no yields are available for 9-11 \ms \ and $>$40 \ms \ 
stars (and since we deliberately neglect yields for intermediate
mass stars), we simply assume that those stars return at their death in the ISM
their initial amount of each element, i.e. that their {\it net yield} 
is zero for all elements (except for deuterium, which is destroyed).
In that way we do not introduce any artificial modification of the
adopted yields.
This procedure is crucial for a correct evaluation of the metal/H ratio at a given
time, especially at late times.

\subsection {Star formation rate and initial mass function}

Observations of average SFR  vs. gas surface density in spirals and starbursts
(Kennicutt 1998) are compatible with a Schmidt type law 
\begin{eqnarray} 
 \Psi(t) \ = \ \nu \ \sigma_{gas}^{k}(t) 
\end{eqnarray}
with ~$k$=1-2. However, this concerns only the {\it disk  averaged} SFR and 
Kennicutt (1998) points out that the local SFR may have a different behaviour.
Indeed, theoretical ideas of SFR in galactic disks suggest a radial dependence
of the SFR (Wyse and Silk 1989) and such a dependence is indeed required in order
to explain the observed abundance, colour and gas profiles in spirals
(Boissier and Prantzos 1999, Prantzos and Boissier 2000). For the purposes of this
work we adopt a Schmidt law with $k$=1.5; when combined with the adopted infall
prescription (see next section) this leads to a slowly varying star formation history
in the galactic disk, compatible with various observables (see Sec. 4).
For consistency, we keep the same form of the SFR in the halo model, although there
is no observational hint for the SFR behaviour during this early stage.

We adopt the IMF  from the work of Kroupa et al. (1993, hereafter KTG93),
where the complex interdependence of several factors  (like stellar
binarity, ages and metallicities, as well as
mass-luminosity and colour-magnitude relationships) is explicitly
taken into account. It is  a three-slope power-law IMF  $\Phi (M)\propto M^{-(1+x)}$;
in the high mass regime it has
a relatively steep slope of $X$=1.7 (based on Scalo 1986),
 while it flattens in the low-mass
range ($X$=1.2 for 0.5$<$M/\ms$<$1. and  $X$=0.3 for M$<$0.5 \ms).
We adopt  this IMF
between 0.1 and 100 \ms, although we are aware that there is some debate as to the
exact form of the low-mass part. Again, for consistency, we adopt the same IMF in the
halo and in the disk model.

\subsection {Gaseous flows: infall and outflow}

In most models of chemical evolution of the solar neighborhood, it is 
implicitly assumed
that the old (halo) and young (disk) stars are parts of the same physical system,
differing only by age;
the same model is used to describe the whole evolution, from the very low
metallicity regime to the current (supersolar) one (e.g. Timmes et al. 1995).

This assumption is, of course, false. The halo and the disk are different 
entities; different processes dominated their evolution, as revealed by the
corresponding metallicity distributions (MD). In the case of the disk,
observations show that the number of metal-poor stars is much smaller
than what is predicted by the simple ``closed-box'' model of chemical
evolution (the ``G-dwarf problem''); 
the simplest explanation of that is that the disk evolved not
as a closed box, but by slowly accreting infalling gas (e.g. Pagel 1997).
In the case of the halo, the observed MD suggests that metal production
was inefficient in those early times; the currently accepted explanation is 
that a strong outflow, at a rate $\sim$9 times the star formation rate,
has occured during the halo evolution (as initially suggested by Hartwick 
1976).

It is clear, then, that a unique model is inadequate to cover the whole
evolution of the solar neighborhood. Still, this is done in most cases.
Only in a handful of works has this point been taken into account, by
adopting different prescriptions for the halo and the disk
(Prantzos et al. 1993, Ferrini et al. 1993, Pardi et al. 1995, 
Chiappini et al. 1997, Travaglio et al. 1999), although not always the
appropriate ones.  The importance of that 
point is twofold: First, the corresponding MDs (the strongest constraints
to the models) are only reproduced when appropriate models are used. 
Secondly, infall and outflow modify the
timescales required for the gas to reach a given metallicity. 
This is important when one is
interested in elements produced by e.g. intermediate mass stars, which enter
late the galactic scene.

Another important point, related to the first one,
is that the halo and the disk are, most probably, not related by any 
temporal sequence. Indeed, the gas leaving the halo ended, quite probably, in 
the bulge of the Galaxy, not in the disk, as argued e.g. by Wyse (2000
and references therein) on the basis of angular momentum conservation
arguments. The disk may well have started with primordial metallicity,
but a very small amount of gas. The corresponding small number of low
metallicity stars that were formed by that gas explains readily
the G-dwarf problem.

In the light of these arguments, we treat then the halo and the disk as
separate systems, not linked by any temporal sequence.
The local disk is assumed to be built up by slow accretion of gas 
with primordial composition. 
An exponentially decreasing infall rate 
 $f(t) \propto e^{-t/\tau}$ with $\tau$ $>$ 7 Gyr
is adopted. Such a long timescale has been shown (Chiappini et al. 1997, 
Prantzos and Silk 1998) 
to provide a satisfactory fit to the data of Wyse and Gilmore (1995) and
Rocha-Pinto and Maciel (1996).
We have normalized the infall rate ~$f(t,R)$, as to obtain the local 
disk surface density $\Sigma_{T}(R)$=55 M$_{\odot}$ pc$^{-2}$ at an age
T=13.5 Gyr. Notice that chemodynamical models also support the idea of 
long time scales for the disk formation (Samland et al 1997).

For the halo model, there are less constraints: neither the duration of the
halo phase, nor the final gas fraction or amount of stars are known. We assume 
then a duration of 1 Gyr and an outflow rate $R_{out}$ = 9 $\Psi(t)$, in order
to reproduce the observed halo MD. For consistency, we use
the same SFR law and the same IMF as in the disk.

\section { Evolution of the halo and the disk}

We run two chemical evolution models, one for the halo (with outflow, 
for 1 Gyr) and one for the disk (with infall, for 13.5 Gyr), 
starting in both cases with gas of primordial composition. The only
observational constraints common for the halo and the disk are:
i) the metallicity distributions of low mass long-lived stars, 
and ii) the element/element ratio vs. metallicity (in particular,
the O vs. Fe evolution). In the case of the disk there are several
more constraints (see Sec. 4.2) but we turn first to (i) and (ii).

\subsection {Metallicity distribution and O vs. Fe in the halo and the disk}

In Fig. 3   we present our results and compare them to observations.
The metallicity distributions ($f$=dN/d[Fe/H]) are normalised 
to $f_{max}$=1
and presented in the upper panel of Fig. 3. The adopted prescriptions
(strong outflow for the halo and slow infall for the disk) lead to
a satisfactory agreement between theory and observations, as expected
on the basis of the discussion in Sec. 3.3. Notice that in the case
of the disk, the theoretical curve shows a low metallicity tail below 
[Fe/H]=-1. However, the number of stars in the tail is extremely
small, less than 10$^{-2}$ of the total. Although there is no
``physical'' discontinuity in the disk population at [Fe/H]=-1, we
systematically show below all our results for the disk corresponding
to [Fe/H]$>$-1 with {\it thick} solid curves, in order to stress that
they correspond to what is traditionally thought as the ``disk phase''
of the Milky Way. Results for [Fe/H]$<$-1 are shown with {\it thin} solid
curves, indicating that such stars do, in principle, exist, but in
very small numbers.

\begin{figure}
\psfig{file=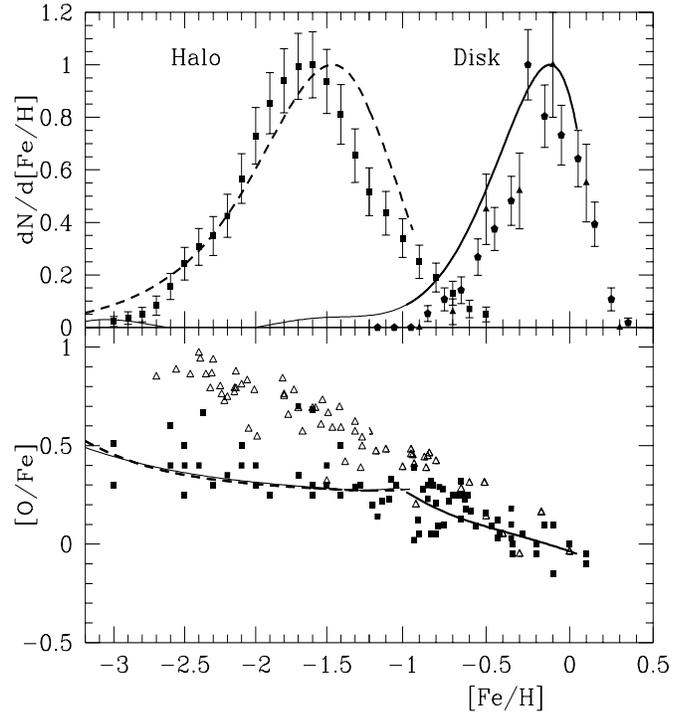,height=10.cm,width=0.5\textwidth}
\caption{\label{} 
{\it Upper panel}: Model metallicity distributions (MD) of the 
galactic halo 
({\it  dashed  curve}) and the local disc ({\it  solid curve}) 
obtained with appropriate models and the metallicity dependant yields  
of WW1995; The ``traditional'' disk population, at [Fe/H]$>$-1, is
indicated by a {\it thick} curve (see Sec. 4.1). 
Observations for halo MD are from Norris and Ryan (1991,
{\it filled squares}) and for the disk from Wyse and Gilmore (1995,
{\it filled pentagons}) and Rocha-Pinto and Maciel 
(1996, {\it filled triangles}). 
{\it Lower panel}: 
[O/Fe] vs. [Fe/H]  in the halo ({\it  dashed curve})
and the disk ({\it solid curve}, {\it thick} for [Fe/H]$>$-1 and
{\it thin} for [Fe/H]$<$-1), according to our model.
Observed abundances are from sources listed in Table 1 ({\it filled squares}),
except for those of Israelian et al. (1998) and Boesgaard 
et al. (1999) ({\it open triangles}). All MDs are normalised to $f_{max}$=1.
}
\end{figure}

Because a large part of Fe in the disk comes from SNIa (at least in our models)
it is not clear whether the final G-dwarf metallicity distribution is mostly
shaped by infall or by the rate of SNIa. In other terms, how can one be
certain that the observed ``G-dwarf problem'' requires indeed large infall
timescales (such as those discussed in Sec. 3.1 and adopted here)?
We notice that the G-dwarf problem concerns mainly the low metallicity regime
i.e. around [Fe/H]=-1 to -0.6; it is in this metallicity range that the closed
box model predicts an excess of low-mass stars w.r.t the observations.
But at those early times,  corresponding to the first $\sim$2-4 Gyr of the
disk's history, the ratio of SNIa/SNII is still small (with the adopted 
prescription for the SNIa rate) and most of the Fe comes from SNII.
Thus, the success of the model in reproducing the G-dwarf metallicity
distribution does rely on the infall prescription, and not on
the SNIa rate prescription. SNIa start becoming major sources of Fe somewhat 
later (around [Fe/H]=-0.5).

In the lower panel of Fig. 3 we show the corresponding evolution of O vs. Fe.
It is virtually identical in the two models, up to [Fe/H]$\sim$-1, since
both elements are primaries and produced in the same site (massive, short-lived,
stars); their abundance ratio is then independant of infall or outflow
prescriptions. As discussed in Sec. 2,
the observed decline of O/Fe in the disk is reproduced by the delayed
appearance of SNIa, producing $\sim$2/3 of the solar Fe.

\begin{figure}
\psfig{file=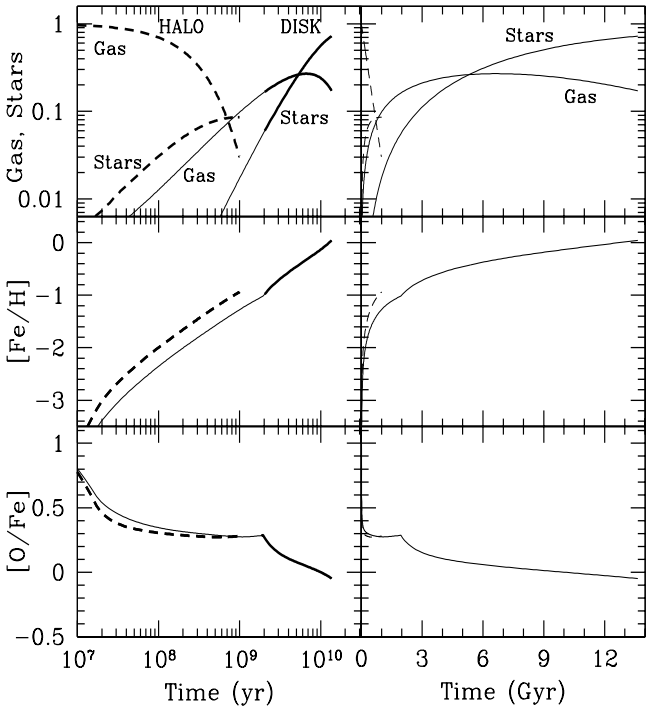,height=10.cm,width=0.5\textwidth}
\caption{\label{} 
Evolution of stars, gas and metals in our models for the halo and the disk, 
plotted as a function of time. A logarithmic time scale is used on the 
{\it left},
in order to show better the halo evolution, whereas the {\it right} panels
are more appropriate for the disk evolution. In all panels, results for the
halo are shown in {\it dashed curves} and for the disk in {\it solid curves}
({\it thick} for [Fe/H]$>$-1 and {\it thin} for [Fe/H]$<$-1).}
\end{figure}

\begin{figure}
\psfig{file=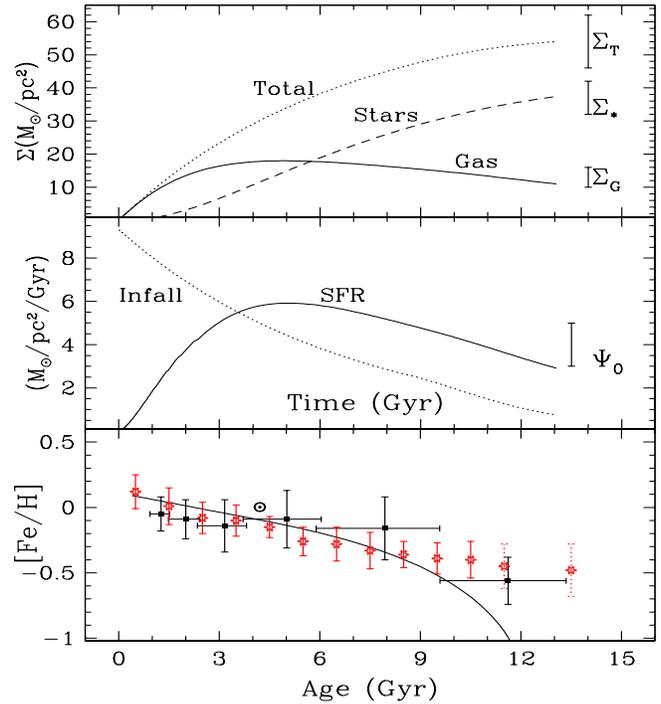,height=10.cm,width=0.5\textwidth}
\caption{\label{} 
Comparison of the  main observables of the solar
neighbourhood to our model predictions. The {\it upper panel}
shows the surface densities of stars, gas and total amount of
matter as a function of time. The vertical error bars represent
present day values. The {\it middle panel} shows the star formation
rate and infall rate; the current SFR ($\Psi_0$) is
indicated by the error bar. 
Data for those tow panels are from the compilation of 
Boissier and Prantzos (1999).
In the  {\it lower panel} the solid curve
shows the derived age-metallicity relation; data are from Edvardsson et al. 
(1993, {\it filled symbols}) and Rocha-Pinto et al. (2000, {\it open 
symbols}, with the last two being rather upper limits), 
while the position of the
Sun is shown by the symbol $\odot$.
}
\end{figure}

Fig. 4 presents the evolution of the halo and the disk in a more
``physical'' way than in Fig. 3, i.e. various quantities are plotted as
a function of time; time is plotted on a logarithmic  scale (on the left,
so that the halo evolution can be followed) and on a linear scale (on
the right). The differences between the two models can be clearly seen.
In particular, at a given time, the metallicity [Fe/H] (middle panel)
is larger in the halo than in the disk (by 0.3 dex, i.e. a factor of 2);
metallicity increases more slowly in our disk model than in the halo one.
It takes $\sim$2 Gyr to the disk to reach [Fe/H]$\sim$-1, compared to 
$\sim$1 Gyr in the case of the halo. However, as noticed already, this
early disk evolution concerns only very few stars.

\subsection {Evolution of the local  disk}

There are many more observational constraints for the local disk than for 
the halo; an extensive presentation of those constraints can be found in 
Boissier and Prantzos (1999, their Table 1 and references therein).
Here we present only briefly those constraints. Besides the MD and the
O vs. Fe evolution, a satisfactory disk model should also reproduce:

(a) The current surface densities of gas ($\Sigma_{G}$), stars
($\Sigma_{*}$), the total amount of matter ($\Sigma_{T}$) and the
current star formation rate ($\Psi_{0}$);

(b) The age-metallicity relationship, traced by the Fe abundance
of long-lived, F-type stars;

(c) The abundances of various elements and isotopes at solar 
birth $(X_{i},\odot)$ and today ($X_{i},0$);

(d) The present day mass function (PDMF), resulting from 
the stellar IMF and the SFR history, which gives an 
important consistency check for the adopted SFR and IMF.

In Fig. 5 we present our results and compare them to constraints (a) and
(b). It can be seen that  the adopted SFR and infall rate lead to a 
current gas surface density of $\Sigma_{G}\sim$ 10 $M_{\odot}$ pc$^{-2}$ 
and a final stellar surface density of
$\Sigma_{*} \sim$36 M$_{\odot}$ pc$^{-2}$, both in good agreement 
with observations.  A current
SFR $\sim$3.5 M$_{\odot}$ pc$^{-2}$ Gyr$^{-1}$ is obtained at
T=13.5 Gyr, also in agreement with observations. The evolution of the
SFR is quite smooth, its current value being about half the
maximum one in the past.

The lower panel of Fig.  5 shows  the disk age-metallicity relation.
The existence of an  age-metallicity relation (AMR) in the disk is one 
of the important issues  in studies of  chemical evolution of the solar
neighborhood.
Several studies in the past  showed a trend of decreasing 
metallicity with increasing stellar ages (Twarog 1980, Meusinger et al. 
1991, and Jonch-Sorensen 1995).  Edvardsson et al (1993) found an 
AMR consistent with these results but with a considerable scatter
about the mean trend.  However, this scatter (difficult to interpret in 
the framework of conventional models), may be due to contamination of the
Edvardsson et al. (1993) sample by stars from different galactic regions 
(Garnett and Kobulnicky 2000). 
Indeed,  the recent survey of Rocha-Pinto et al (2000, also on Fig. 5), 
suggests a scatter almost half of that in Edvardsson et al. (1993).
In view of the
current uncertainty, we consider that the mean trend of the disk AMR
obtained with our model is in satisfactory agreement with observations.  

In Fig. 6 we compare  our results to constrain (c), i.e to the
elemental (upper panel) and isotopic (lower panel) composition of the Sun.
It is assumed that the Sun's (and solar system's) composition is 
representative of the one of the local interstellar medium (ISM) 4.5
Gyr ago, but this assumption is far from been definitely established. 
Indeed,   CNO abundances in young
stars and gas in the nearby Orion nebula show that the metallicity of this
young region is lower than solar (Cunha \& Lambert 1994,
Cardelli \& Federmann 1997); this cannot be readily
interpreted in conventional models of chemical evolution.
On the other hand, the Fe abundance of young
stars determined by Edvardsson et al. (1993) seems to be compatible with
the conventional picture, while the data of Rocha-Pinto et al (2000)
suggest that the Sun is indeed Fe-rich w.r.t. other stars of similar age
(Fig. 5). One should keep in mind this question (of the Sun
being ``typical'' or not) when making detailed comparison of its
composition to model predictions.

\begin{figure*}
\psfig{file=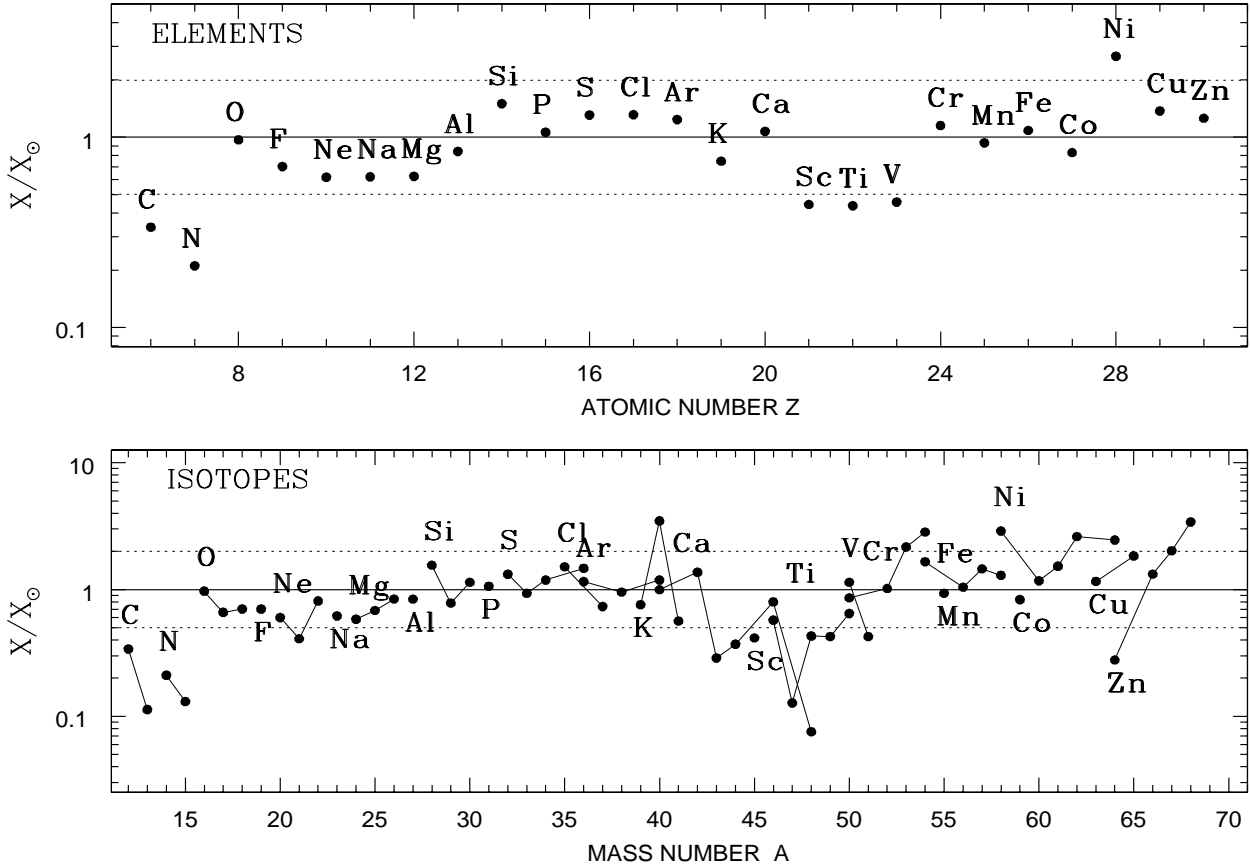,angle=-90,height=13.cm,width=\textwidth}
\caption{\label{} 
Ratio of the calculated and observed solar abundances of
elements C to Zn ({\it upper panel}) and their stable isotopes 
({\it lower panel}).
Results of our model are shown at a disk age of 8.5 Gyr
(Sun's formation), and yields
from massive stars (metallicity dependant, from WW1995) and SNIa (from
the W7 model of Thielemann et al. 1986) are taken into account.
The {\it dotted lines} mark deviations by a factor of 2 from the solar
composition.
All currently available sets of massive star yields show an underproduction
of Sc, Ti and V. C and N also require additional production sources.
The overproduction of Ni (in the form of the main isotope $^{58}$Ni)
results from the W7 model of Thielemann et al.  (1986) for SNIa. }
\end{figure*}

An inspection of Fig. 6 shows that there is good overall 
agreement between theory  and observations, i.e. about 80\% of the elements
and isotopes are co-produced within a factor of two of their solar
values. One should notice the following:

-  The carbon isotopes require another source. $^{12}$C may be produced
either by intermediate mass stars, as usually assumed, or by Wolf-Rayet
stars with metallicity dependant yields (Maeder 1992, Prantzos et al. 1994).
$^{13}$C is made probably in intermediate mass stars. The evolution of 
$^{12}$C/$^{13}$C in the disk and its implications for the synthesis of
those isotopes is studied in Prantzos et al. (1996).

- The nitrogen isotopes also require another source. $^{14}$N has the same 
candidate sites as $^{12}$C. Novae seem to be a viable source for $^{15}$N,
but current uncertainties of nova models do not allow definite
conclusions.

- Fluorine is produced by neutrino-induced nucleosynthesis in WW1995, and
this is also the case for a few other rare isotopes, not shown in Fig. 6
($^7$Li, $^{11}$B). This is an interesting ``new'' nucleosynthesis
mechanism, but because of the many involved uncertainties (see Woosley et
al. 1990) it cannot be considered as established yet. One should keep an 
``open eye'' for other, more conventional,  sites of fluorine (as well 
as lithium and boron) nucleosynthesis, like e.g. Wolf-Rayet stars
(Meynet and Arnould 1999).

- The obtained overabundance of $^{40}$K may reflect the large uncertainty
in the abundance of this isotope at solar system formation (see
Anders and Grevesse 1989), as already pointed out in Timmes et al. (1995). 

- Sc, V and Ti isotopes are underproduced, indicating that all
currently available models of massive stars have some problems with the
synthesis of these species.

- There is a small overproduction of Ni, 
due to the isotope $^{58}$Ni, which is abundantly produced
in the W7 and W70 models of SNIa. This is also true for $^{54}$Cr, 
a minor isotope of Cr. The amount of those nuclides depends mostly 
on the central density of the exploding white dwarf and the 
overproduction problem may be fixed by varying this parameter.
Alternatives to the W7 model
have recently been calculated by Iwamoto et al. (1999). On the other hand,
Brachwitz et al. (2000) have explored the effect of electron capture rates
on the yields of Chandrasekhar mass models for SNIa; they showed that
the problem with $^{54}$Cr may disappear (depending on the ignition density)
while the one with  $^{58}$Ni is
slightly  alleviated. It can be reasonably expected that in future,
improved, SNIa models, the overproduction problem of those nuclei will be
completely solved.

Notice that in our calculation, the Fe-peak isotopic yields
of WW1995 have been reduced by a factor of two, in order to reproduce
the observed O/Fe ratio in halo stars ($\sim$3 times solar, see Fig. 3
and Sec. 5);  otherwise, the WW1995 massive
stars alone can make almost the full solar abundance of Fe-peak nuclei (as
shown in Timmes et al. 1995), leaving no room for SNIa. Taking into account
the uncertainties in the yields, especially those of Fe-peak nuclei (see
Sec. 2) our reduction imposed on the WW1995 Fe yields is not unrealistic.

The nice agreement between theory and observations in Fig. 6 comes as
a pleasant surprise, in view of the many uncertainties discussed in the
previous section. It certainly does not guarantee that each individual
yield is correctly evaluated. It rather suggests that the various factors
of uncertainty  cancel out (indeed, it is improbable that they all
``push'' towards the same direction!)  
so that a good  overall agreement with observations results. 
As stressed in Timmes et al. (1995), the abundances of the
intermediate mass isotopes span a range of 8 orders of magnitude; reproducing 
them within a factor of two suggests that models of massive stars nucleosynthesis
are, globally, satisfactory.
At least to first order, 
currently available yields of massive stars + SNIa can indeed account for the solar 
composition between O and Zn (with the exceptions of Sc, Ti and V, and possibly F).

\begin{table*}[h]
{\scriptsize
{\bf TABLE 1: Reference list of the observational data for the halo and the disk stars}\\

\begin{tabular}{ c c c c c c c c c c c c c c c c c c c c c}
\hline \\
 C& N& O& Na& Mg& Al& Si& S&  K& Ca& Sc& Ti& V& Cr& Mn& Fe& Co& Ni& Cu& Zn& Ref \\   
  &  &  &   &   &   &   &  &   &   &   &   &  &   &   &   &   &   &   &    &      \\
\hline
x &  &  &   &   &   &   &  &   &   &   &   &  &   &   &  x&   &   &   &    &  1   \\
  &  &  &   & x & x &   &  &   &   &   &   &  &   &   &  x&   &   &   &    &  2  \\
  &  &  & x & x & x & x &  &   & x & x & x & x& x & x &  x& x & x & x &    &  3    \\
x & x&  &   &   &   &   & x&   &   &   &   &  &   &   &  x&   &   &   &    &  4  \\
  &  &  &   &   &   & x &  &   &   &   & x &  &   &   &  x&   & x &   &    &  5  \\
x & x&  &   & x & x &   &  &   & x &   & x &  &   &   &  x&   & x &   &    &  6  \\
x & x&  &   &   &   &   &  &   &   &   &   &  &   &   &  x&   &   &   &    &  7  \\
x &  & x& x & x & x & x &  &   & x &   & x &  &   &   &  x&   & x &   &    &  8   \\
  &  &  & x & x & x & x &  &   & x & x &   &  &   &   &  x&   &   &   &    &  9   \\
  &  &  & x & x & x & x &  &   &   &   &   &  &   &   &  x&   &   &   &    &  10   \\
  &  &  & x & x &   &   &  &   &   &   &   &  &   &   &  x&   &   &   &    &  11  \\
x & x& x&   &   &   &   &  &   &   & x &   &  &   &   &  x&   &   &   &    &  12  \\
  &  &  &   & x &   &   & x&   &   &   &   &  &   &   &  x&   &   &   &    &  13  \\
x & x&  &   &   &   &   &  &   &   &   &   &  &   &   &  x&   &   &   &    &  14  \\
  &  &  &   &   &   & x &  &  x& x &   & x &  &   &   &  x&   & x &   &    &  15 \\
  &  &  & x & x & x & x &  &   & x &   &   &  &   &   &  x&   &   &   &    &  16 \\
  &  &  &   &   &   &   & x&   &   &   &   &  &   &   &  x&   &   &   &    &  17 \\
  &  &  &   &   &   &   &  &   & x & x & x & x& x & x &  x& x & x &   &    &  18  \\
  &  &  & x & x & x & x &  &   & x & x & x &  & x &   &  x&   & x &   &    &  19 \\
  &  &  &   & x & x &   &  &   & x & x & x &  & x &  x&  x&   & x &   &    &  20 \\
  &  &  &   &   &   &   &  &   &   &   &   &  &   &   &  x&   & x &   &  x &  21 \\
  &  &  &   &   &   &   &  &   &   &   &   &  &   &  x&  x&   &   &   &    &  22 \\
  &  &  &   & x & x &   &  &   & x & x & x &  & x &   &  x&   &   &   &    &  23 \\
  &  &  &   & x & x & x &  &   & x &   & x &  &   &   &  x&   & x &   &    &  24 \\
  &  &  & x & x &   &   &  &   & x &   & x &  &   &   &  x&   &   &   &    &  25 \\
  &  &  & x &   &   &   &  &   & x & x & x &  & x &   &  x&   & x &   &    &  26  \\
  &  & x&   &   &   &   &  &   &   &   &   &  &   &   &  x&   &   &   &    &  27 \\
  &  &  &   &   &   & x &  &   & x & x & x & x& x &  x&  x& x & x &   &    &  28 \\
  &  &  &   & x & x & x &  &   & x & x & x &  & x &  x&  x& x & x &   &    &  29 \\
  &  &  &   &   &   &   &  &   &   &   &   &  &   &   &  x&   & x &   &  x &  30 \\
  &  & x&   &   &   &   &  &   &   &   &   &  &   &   &  x&   &   &   &    &  31 \\
  &  & x&   &   &   &   &  &   &   &   &   &  &   &   &  x&   &   &   &    &  32 \\
  &  & x&   &   &   &   &  &   & x &   &   &  &   &   &  x&   &   &   &    &  33 \\
x &  &  &   &   &   &   &  &   &   &   &   &  &   &   &  x&   &   &   &    &  34 \\
  &  & x& x & x & x & x &  &   & x &   & x &  &   &   &  x&   & x &   &    &  35  \\
  &  &  &   & x & x &   &  &   & x & x & x &  & x &  x&  x&   & x &   &    &  36\\
x &  &  &   &   &   &   &  &   &   &   &   &  &   &   &  x&   &   &   &    &  37  \\
x & x& x& x & x & x & x &  &   & x & x & x & x& x &  x&  x& x & x &   &  x &  38 \\
x & x& x&   &   &   & x &  &   &   &   &   &  &   &   &  x&   &   &   &    &  39 \\
  &  & x&   &   &   &   &  &   &   &   &   &  &   &   &  x&   &   &   &    &  40 \\
  &  & x&   & x &   &   &  &   & x &   & x &  & x &   &  x&   &   &   &    &  41 \\
  &  &  & x & x & x & x &  &   & x &   & x &  &   &   &  x&   &   &   &    &  42 \\
x &  &  & x & x & x & x &  &   & x & x & x &  & x &  x&  x& x &   &   &    &  43 \\
  &  &  &   & x &   &   &  &   &   &   &   &  &   &   &  x&   &   &   &    &  44 \\
  &  & x&   &   &   &   &  &   &   &   &   &  &   &   &  x&   &   &   &    &  45 \\
x &  &  & x & x & x & x &  &   & x & x & x & x& x &  x&  x& x & x &   &    &  46  \\
x &  &  &   &   &   &   &  &   &   &   &   &  &   &   &  x&   &   &   &    &  47  \\
x &  & x&   &   &   &   &  &   &   &   &   &  &   &   &  x&   &   &   &    &  48 \\
  &  &  &   & x & x & x &  &   &x  &x  & x &  & x & x &  x& x & x &   &    &  49 \\  
  &  & x& x & x &   & x &  &   & x &   & x &  & x &   &  x&   & x &   &    &  50 \\
x &  & x& x & x &   &   &  &   & x & x & x &  & x &   &  x&   &   &   &    &  51 \\
  &  &  &   &   & x &   &  &   &   &   &   &  &   &   &  x&   &   &   &    &  52  \\
  &  & x&   &   &   &   &  &   &   &   &   &  &   &   &  x&   &   &   &    &  53 \\
  &  & x& x & x & x & x &  &   & x & x & x & x& x & x &  x& x & x &   &    &  54   \\
  &  &  &   & x &   &   &  &   & x &   & x & x& x &   &  x&   & x &   &    &  55 \\
  &  & x&   &   &   &   &  &   &   &   &   &  &   &   &  x&   &   &   &    &  56 \\
  &  &  &   &   &   &   &  &   &   & x &   &  &   &  x&  x&   &   &   &    &  57 \\
  &  & x& x & x & x & x &  & x & x &   & x & x& x &   &  x&   & x &   &    &  58  \\
  &  &  & x & x &   & x &  &   & x &   & x &  & x &   &  x&   & x &   &    &  59  \\
x & x& x& x & x &   &   &  &   &   &   &   &  &   &   &  x&   &   &   &    &  60  \\
\hline
\end{tabular}

References:
1. Sneden et al. 1979 ;
2. Carney \& Peterson 1981  ;    
3. Peterson 1981;
4. Clegg et al. 1981     ;
5. Leep \& Wallerstein 1981   ;
6. Barbuy et al. 1985   ;
7. Laird 1985    ;
8. Magain 1985   ;
9. Tomkin et al. 1985 ;
10.Francois 1986a;
11.Francois 1986b;
12. Gratton \& Ortolani 1986;
13. Francois 1987a;
14. Carbon et al. 1987;
15. Gratton \& Sneden 1987;
16. Magain 1987;
17. Francois 1987b;
18. Gilroy et al. 1988;
19. Gratton \& Sneden 1988;
20. Hartmann \& Gehren 1988;
21. Sneden \& Crocker 1988;
22. Gratton 1989;
23. Magain 1989;
24. Molaro \& Castelli 1990;
25. Molaro \& Bonifacio 1990
26. Zhao \& Magain 1990;
27. Bessel et al. 1991;
28. Gratton \& Sneden 1991;
29. Ryan et al. 1991;
30. Sneden et al. 1991;
31. Spite \& Spite 1991;
32. Spiesman \& Wallerstein 1991;
33. Nissen \& Edvardsson 1992;
34. Tomkin et al. 1992;
35. Edvardsson et al. 1993;
36. Norris et al. 1993;
37. Andersson \& Edvardsson 1994;
38. Beveridge \& Sneden 1994
39. Cunha \& Lambert 1994;
40. King 1994;
41. Nissen et al. 1994;
42. Primas et al. 1994;
43. Sneden et al. 1994;
44. Fuhrmann et al. 1995;
45. King \& Boesgaard 1995;
46. McWilliam et al. 1995;
47. Tomkin et al. 1995;
48. Balachandran \& Carney 1996;
49. Ryan et al. 1996;
50. Nissen \& Schuster 1997;
51. Baum\"uller \& Gehren 1997;
52. Laimons et al. 1998;
53. Israelian et al. 1998;
54. Feltzing \& Gustafsson 1998;
55. Jehin et al. 1999;
56. Boesgaard et al. 1999;
57. Nissen et al. 1999;
58. Chen et al. 1999;
59. Stephens A. 1999;
60. Carretta et al. 2000.
}
\end{table*}

\section {Abundance ratios in the halo and the  local disk}

We calculated the abundance evolution of all the isotopes between 
H and Zn in the framework of our halo and local disk models, by
using two different sets of massive star yields: i) the yields
of WW1995 at constant (=solar) metallicity (Case A in the following), 
and ii) the metallicity dependant yields of WW1995, by interpolating
between the values given for metallicities Z/Z$_{\odot}$=0, 10$^{-4}$,
10$^{-2}$, 10$^{-1}$ and 1 (Case B in the following). 
Because of our neglect of the C and N yields of intermediate mass stars,
total metallicity is not consistently calculated in our models; we use
oxygen as metallicity indicator, in order to inerpolate in the WW1995 
tables (in the WW1995 models, the initial abundances of all elements
are scaled to metallicity). Obviously,
Case B (also studied by Timmes et al. 1995) is the ``reference''
case, whereas Case A is only studied for illustration purposes.  
In both cases, the yields of the W7 and W70  models of Iwamoto et al (1999)
for SNIa are used (interpolated as a function of metallicity),
whereas no yields from intermediate mass stars
are considered; our explicit purpose is to check to what extent
massive stars can account for observations of intermediate mass elements
and for which elements the contribution of intermediate mass stars 
is mandatory. We stress again that  we {\it do take into account the
contibution of intermediate and low mas stars to the H and He ``budget''},
since this is crucial for a correct evaluation of the metal/H ratio,
especially at late times (Sec. 3.1).

Since most of the available data on the composition of stars concerns
elemental abundances, we computed the corresponding evolution by
summing over the calculated isotopic abundances. We present our results
in Fig. 7 and compare them to a large body of observational data; 
most of the data come from the surveys of Ryan et al. (1996) and Mc William
(1997) for the halo and Edvardsson et al. (1993) and Chen et al. (2000)
for the disk, but we
included a large number of other works, concerning specific elements
(the corresponding references are listed in Table 1). 
We do not attempt here any discussion on the quality of these data (this
would be beyond the scope of this work), 
and we refer the reader to the recent review of Ryan (2000) for that.
It is obvious that systematic
differences between various studies introduce a scatter larger than the
real one (and, perhaps, unrealistic trends in some cases).
Our reference Case B
is shown in {\it thick curves} ({\it dashed} for the halo and {\it solid} 
for the disk), while Case A is in {\it thin curves}.

Before presenting our results we notice that in our models
metallicity reaches [Fe/H]$\sim$-4 at a time t$\sim$10$^7$ yr and
[Fe/H]$\sim$-3 at a time t$\sim$2 10$^7$ yr; these timescales correspond
to the lifetimes of stars of mass M$\sim$20 M$_{\odot}$ and 
 M$\sim$10 M$_{\odot}$, respectively. Any variations in the abundance
ratios in the metallicity range -4 $<$ [Fe/H] $<$ -3
results then from the fact that stars of different  masses
(starting from 100 M$_{\odot}$ and going to 10  M$_{\odot}$)
enter progressively the galactic scene. The discussion of Sec. 2 shows that the
yields of individual stars are very uncertain, much more than those
integrated over the IMF (the latter reproduce, at least, the
solar composition!). Besides, there is absolutely no guarantee
that the model reproduces correctly the relation between age 
and metallicity at those early times. 
For instance, in a recent work Argast et al. (2000) find that the halo
became chemically homogeneous and reached [Fe/H]=-3 after $\sim$160 Myr,
a duration six times longer than in our calculations.
For those reasons we consider that
any abundance trends of our models at [Fe/H]$<$-3 {\it are not
significant}, but we show them for completeness. Integration over the
whole IMF of massive stars is only made for [Fe/H]$>$-3 and we consider
that our results are significant only after that point. 
Finally, we notice that we have reduced the WW1995 yields of Fe-peak isotopes 
by a factor of two, in order to reproduce the observed $\alpha$/Fe
ratio in the halo.

\subsection {Carbon and Nitrogen}

Observations indicate a flat [C/Fe]$\sim$0 in the halo and the disk, with
a large dispersion at all metallicities. Both our cases A and B show indeed
[C/Fe]$\sim$0 in the halo (since both C and Fe are primaries), and a slow
decline of C/Fe in the disk due to Fe production by SNIa. As discussed in
Sec. 2, a complementary source of C is required in the disk. This may
be either intermediate mass stars (IMS) or Wolf-Rayet (WR) stars.
However, as discussed in Prantzos et al. (1994), IMS have masses
M$>$3 M$_{\odot}$ and lifetimes $\tau<$5 10$^8$ yr. Such stars can
certainly evolve during the halo phase ({\it if} the duration of that phase
is indeed $\sim$1 Gyr, as assumed here) and enrich the halo with
C, thus rising the C/Fe ratio at [Fe/H]$<$-1. Such a behaviour is
not observed, however, suggesting either that low mass stars 
(M$<$2 M$_{\odot}$) or WR stars are the main carbon sources in the disk.
The latter possibility is favoured in Prantzos et al. (1994) and Gustafsson
et al. (1999)

\begin{figure*}
\psfig{file=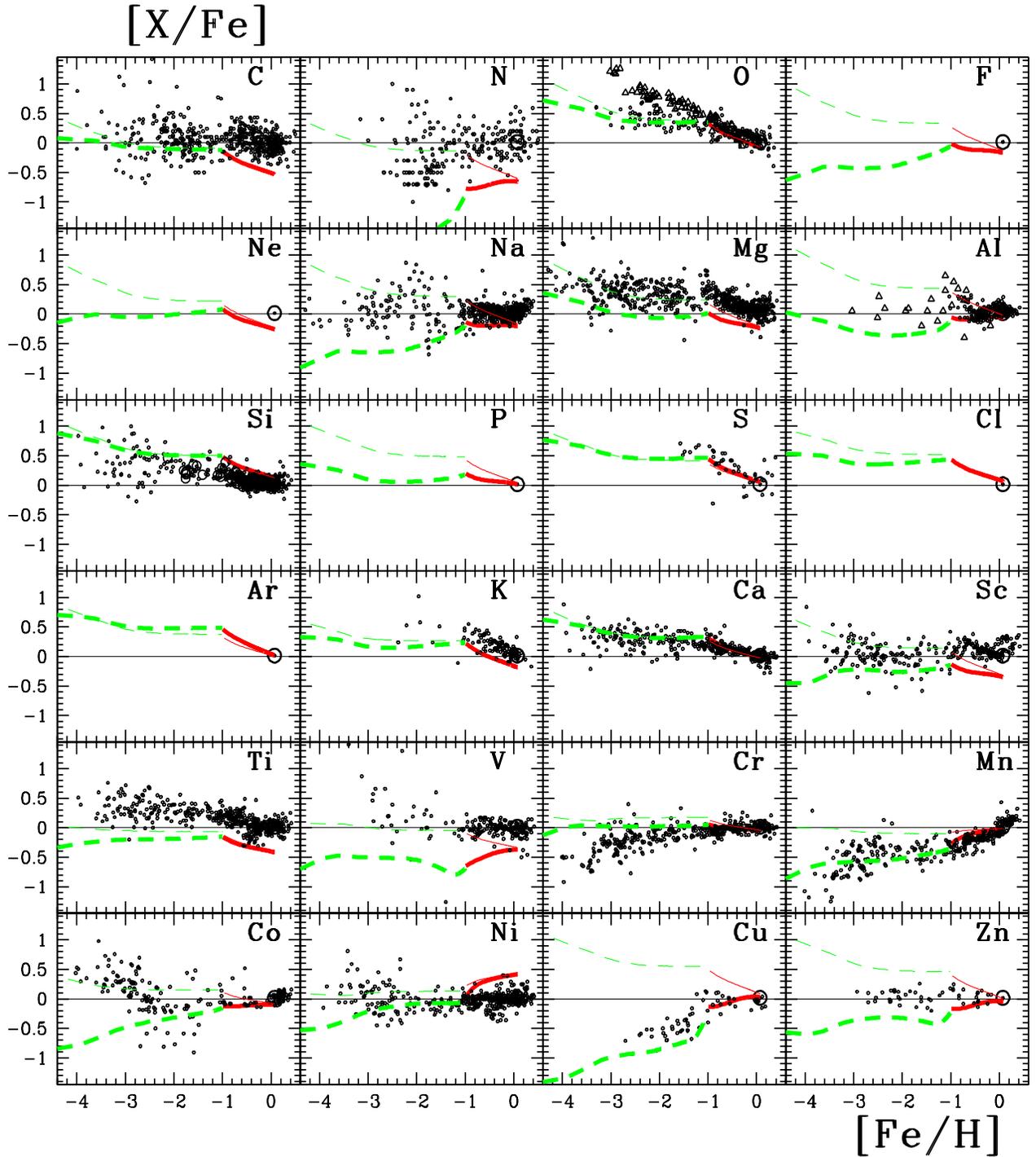,height=20.cm,width=1.05\textwidth}
\vspace{-0.6cm}
\caption{\label{} 
Abundance ratios [X/Fe] of stars in the halo and the local disk, as a
function of [Fe/H]. Theoretical  results are obtained
with  models that treat properly the halo ( {\it dashed curve}  
assuming {\it outflow}) and the disk ( {\it solid curve}  assuming 
{\it slow infall}). Two sets of massive star yields are used, both from WW1995:
at constant (=solar) metallicity ({\it thin curves}, Case A,
only for illustration
purposes) and at variable metallicity ({\it thick curves}, the reference
Case B).  Yields of the W7 and W70 models of Iwamoto et al. (1999) for SNIa
are used in both cases (properly interpolated as a function of metallicity);
intermediate mass stars are not considered.
It should be noted that WW1995 yields of Fe have been divided by 2, in order to 
obtain the observed ${\alpha}/Fe$ ratio in halo  stars. Model trends below 
[Fe/H]=-3 are due to the finite lifetime 
of stars ([Fe/H]=-4 is attained at 10 Myr, corresponding to the lifetime
of stars with mass $> 20 M_{\odot}$, while [Fe/H]=-3 is attained at 20 Myr,
corresponding to the lifetime of $\sim 10 M_{\odot}$ stars). In view 
of the yield  uncertainties of individual stars (Sec. 2) and of the
uncertainties in the timescales at those early times of the halo
evolution, those trends {\it should not be considered as significant}.
The observed data points in the figure are taken from sources listed 
in  Table 1. 
Observed abundance ratios of [O/Fe] from Israelian et al (1998) and 
Boesgaard et al (1999) are shown by {\it open triangles}; they suggest a trend
quite different from all other alpha-elements.
The {\it open triangles} in the [Al/Fe] panel
correspond to observed data with NLTE corrections (from Baum\"uller \& Gehren
1997).}
\end{figure*}

Nitrogen behaves in a similar way as carbon, i.e. the observed [N/Fe]$\sim$0
in the halo and the disk, with a large scatter at low metallisities. Our Case
A (metallicity independant yields) shows also a flat [N/Fe]$\sim$0 evolution
in the halo and a decline in the disk, exactly as for carbon. However, in
the realistic Case B, N behaves as secondary: [N/Fe] increases steadily
up to [Fe/H]$\sim$-1. Its value remains $\sim$constant in the disk phase,
because Fe production by SNIa compensates for the larger N yields of more
metal rich stars. However, the final N/Fe is only $\sim$1/3 its solar value.

Obviously, curent massive star yields fail, qualitatively and quantitavely,
to reproduce the observed evolution of N/Fe. What are the alternatives?
in our view, there are two:

a) Intermediate mass stars, producing primary N through hot-bottom burning
in the AGB phase,
are the most often quoted candidate. Large uncertainties still affect that
complex phase of stellar evolution, but recent studies (e.g. Lattanzio 1998
and references therein) find that
hot-bottom burning does indeed take place in such stars.
If N is indeed produced as a primary in IMS, and their N yields are
metallicity independant, then the N/Fe in the disk should decline
(because of SNIa). Metallicity dependant N yields  from WR stars (Maeder 1992)
could compensate for that, keeping the N/Fe ratio $\sim$constant in the
disk. On the other hand, if N from massive stars is indeed secondary,
at some very low metallicity level (let's say
[Fe/H]$<$-3) the N/Fe ratio should also decline; this would be an important 
test of IMS being the main N source in the halo. If such a decline is
not observed, we are lead to the second alternative, namely

b) Massive stars, producing primary N by an as yet unidentified mechanism,
obviously requiring proton mixing in  He-burning zones.
Such mixing does not occur in standard stellar models, but ``new generation''
models including rotation offer just such a possibility (Heger et al. 1999,
Maeder and Meynet 2000).
In that case, N is produced not by the original
carbon entering the star, but by the carbon produced in He-burning; as
a consequence, it is produced as a primary. In that case, massive stars
could be the main  source of N and C in the halo.

The discussion of this section suggests then an intriguing possibility:
massive stars could well be the main source of C and N in both the halo
and the disk (in the latter case, through the WR winds), leaving only a
minor role to intermediate mass stars!

\subsection {$ \alpha$ - elements ~O, ~Mg, ~Si,~S, ~Ca, ~Ti}

The alpha elements (O, Mg, Si, S, Ca, Ti) present a well known behaviour. The
$\alpha$/Fe ratio is $\sim$constant in the halo, at 
[$\alpha$/Fe]$\sim$0.3-0.5 dex, and declines gradually in the disk. The latter
feature is interpreted as due to (and constitutes the main evidence for) the
contribution of SNIa to the disk composition.

This behaviour is indeed apparent in Fig. 7; despite the large scatter, all the
alpha elements show the aforementioned trend. We stress here again that the 
recent data of Israelian  et al. (1998) and Boesgard et al. (1999),
also plotted in Fig. 7 (with different symbols), challenge this
picture in the case of oxygen. If true, these new data should impose some revision
of our ideas on massive star nucleosynthesis, probably along the lines 
suggested in Sec. 6.

Until the situation is clarified, we stick to the ``old paradigm''.
In the framework of this ``paradigm'', 
Pagel and Tautvaisiene (1995) have shown that the $\alpha$/Fe evolution can
be readily explained by a very simple model (with IRA), the
metallicity independant yields of Thielemann et al. (1996) and SNIa during the
disk phase. On the other hand, Timmes et al. (1995), using the metallicity
dependant yields of WW1995 (but an inappropriate model for the halo, see Sec. 3.3), 
found good agreement with observations, provided that the Fe yields of WW1995 are
reduced by a factor of $\sim$2.

Our results in Fig. 7 point to the following:

- For O, Si, S and Ca, both Cases A and B give virtually identical results. These
elements behave as true primaries, without any metallicity dependence of their
yields. Moreover, after the WW1995 Fe yields are reduced by a factor of 2, a
fairly good agreement with observations is obtained.

- The situation is far less satisfactory for Mg and Ti. For both of them,
the WW1995 yields at solar metallicity are larger than at lower metallicities
(see Fig. 1). This is puzzling since Mg and Ti are also supposed to be primaries
(in fact, more puzzling in the case of Mg, since Ti is produced close to the
``mass-cut'' and subject to more important uncertainties). As a result, our Case
A is marginally compatible with observations of Mg/Fe; the reference Case B
does not match at all the observations, despite the reduction of the Fe yields
by a factor of 2. In the case of Ti, both Cases A and B fail to match the
observations.

These features were also noticed in Timmes et al. (1995) and the problem with
the WW1995 yields of Mg and Ti pointed out; however, no satisfactory alternative was
suggested. 
Since the Mg yields of WW1995 are steeply increasing function of stellar mass,
our use of the Kroupa et al. (1993) IMF 
(steeper than the Salpeter IMF used by Timmes et al. 1995) leads to a low
Mg/Fe ratio, even after reduction of the Fe yields.
Our Fig. 1 (lower panel) suggests that the yields of LSC2000 could
match better the halo data, since   the Mg/Fe and Ti/Fe ratios obtained for
Z=0 are larger than solar. On the other hand, Fig. 1 shows that in both WW1995 
and LSC2000, Mg and Ti have lower overproduction factors than all the other alpha
elements, at all metallicities; this means that, even if the halo Mg/Fe and Ti/Fe
ratios are better reproduced with the LSC2000 yields, the corresponding 
$\alpha$/Mg and  $\alpha$/Ti ratios will certainly not match the observational 
data. Thus, at present, none of the two available sets of metallicity dependant 
yields offers a solution to the problem of Mg and Ti. 

The fact that Pagel and Tautvaisiene (1995) find good agreement with observations
by using the Thielemann et al. (1996) yields may suggest that this set of yields
indeed solves the problem. 
This is also the case in Chiappini et al. (1999), who use a somewhat different 
prescription for SNIa rate than here, and metallicity independant yields
from Thielemann et al. (1996) and WW19995.
Notice, however, that metallicity independant yields (those of Thielemann et al. 1996
are for solar metallicity only) should not be used for studies of the
halo, even if the problem is less severe in the case
of primary elements. The equivalent set of WW1995 yields
also reproduces the Mg/Fe evolution in the halo (our Case A), but it is not 
appropriate. We need to understand how massive stars make a $\sim$constant 
 Mg/Fe and Ti/Fe ratio at all metallicities, by using  stellar models
with the appropriate initial metallicity.

\subsection {Sodium and Aluminium}

Na and Al are two monoisotopic, odd elements. Their theoretical yields
are, in principle, affected by the ``odd-even'' effect (see Sec. 2). This
effect seems to be stronger in the case of LSC2000 than in WW1995 (Fig. 1), at
least for the adopted IMF.

The observational situation for those elements  is not quite clear.
Recent observations (Stephens 1999) suggest that Na/Fe decreases as
one goes from [Fe/H]=-1 to [Fe/H]=-2, as expected theoretically. However,
most other observations do not support this picture,  showing instead
a flat [Na/Fe]$\sim$0 ratio with a large scatter.
Our Case A evolution of Na/Fe is similar to the $\alpha$/Fe evolution
and, obviously, incorrect. In Case B, Na/Fe increases steadily after
[Fe/H]$\sim$-2.5 and reaches a plateau after [Fe/H]$\sim$-1. Neither
case matches the observations well. As we shall see in Sec. 5.7, the
situation improves  considerably when only the halo 
data of Stephens (1999) and the disk data of Edvardsson et al. (1993) 
and Feltzing and Gustafsson (1998) are used; then Na vs. Ca shows the behaviour
of an odd element, as it should.  

Ryan et al. (1996) find a steep decline of Al/Fe at low metallicities, 
down to ``plateau'' value of [Al/Fe]$\sim$-0.8,
but they stress that their analysis neglects NLTE effects and underestimate
the real Al/Fe ratio;  for that reason we do not plot their data in Fig. 7
(Ryan et al. 1996 suggest that a NLTE correction to their data would move
the ``plateau'' value to [Al/Fe]$\sim$-0.3, i.e. consistent with
what expected for an odd-Z element). On the other hand,
the NLTE analysis of the data of Baum\"uller and Gehren (1997, {\it open
triangles} in Fig. 7) 
suggests a practically flat Al/Fe ratio in the halo, a rather unexpected
behaviour for an ``odd'' element.
In our model Case A, Al behaves like an $\alpha$ element.
In Case B, the ``odd-even'' behaviour is manifest: a small increase of 
Al/Fe is obtained as metallicity increases 
from [Fe/H]$\sim$-2.5 to [Fe/H]$\sim$-1 (the model trend 
below [Fe/H]=-3, due to stellar  mass and lifetime effects, is not
significant, as stressed in the begining of Sec. 5).  Once
again, theory does not match observations and observations do not show 
the expected behaviour.

It should be noted at that point that intermediate mass stars of low 
metallicity could, perhaps, produce some Na and Al through the 
operation of the Ne-Na and Mg-Al cycles in their H-burning shells
and eject them in the interstellar medium through their winds.
There are indeed, indications, that in low mass, low metallicity stars
of globular clusters such nucleosynthesis does take place 
(Kraft et al. 1998). If this turns out to be true also for intermediate
mass stars of low metallicity, it might considerably modify our ideas 
of Na and Al nucleosynthesis in the halo.

\subsection {Potassium, Scandium, Vanadium }

K, Sc and V are three odd-Z elements produced mainly by oxygen
burning. However, the first one is produced in hydrostatic burning
and the other two in explosive burning, i.e. their nucleosynthesis is more
uncertain. Their yields are affected in  similar ways by the initial
metallicity of the star, as can be seen in Fig. 1. 

Currently available 
observations show a rather different behaviour  for those elements:
Sc/Fe remains $\sim$solar in the whole metallicity range -3$<$[Fe/H]$<$0.
V/Fe is also $\sim$solar in the disk and the late halo, but  
appears to be supersolar in the range $-3<$[Fe/H]$<$-2 (although the data is 
rather scarce for a definite conclusion). Finally, K/Fe declines in the disk,
while the rare halo data point to supersolar ratio [K/Fe]$\sim$0.5, i.e.
its overall behaviour is similar to that of an $\alpha$-element!

From the theoretical point of view, the situation is also unsatisfactory.
Cases A and B produce distinctively different results for Sc and V, but not
so for K. In Case B, the Sc/Fe and V/Fe ratios are subsolar in the halo, 
while K/Fe is supersolar. Also, in that case, K/Fe declines in the disk, 
Sc/Fe remains $\sim$constant and V/Fe increases.

This ``strange'' theoretical behaviour results from the interplay of several
factors, which do not affect all those elements in the same way:
odd-even effect, Fe yield reduction and contribution of SNIa.
Thus, the metallicity dependence of the yields between Z=0.1 Z$_{\odot}$
and Z=Z$_{\odot}$ is stronger for V than for the other two. In fact, the V yield
at metallicity Z=0.1 Z$_{\odot}$ is lower than at Z=0.01 Z$_{\odot}$ in
WW1995, which is counterintuitive (making V/Fe to decrease between
[Fe/H]=-2 and [Fe/H]=-1).
Also, SNIa contribute more to the production of V than to the one of Sc
or K (at least according to the W7 model). For those reasons, Sc/Fe is
$\sim$ constant in the disk, while K/Fe declines and V/Fe increases.

Although our Case B seems to match well the available data for K, we think
that this is rather fortuitous: we obtain a supersolar K/Fe in the halo
because of the reduction of the Fe yields by a factor of two and 
of the adopted IMF (Timmes et al. 1995 obtain a solar K/Fe in the halo
for the same reduced Fe yields, probably because they use the Salpeter IMF).

In our view, the evolution of those three elements is far from being well
understood, either observationaly or theoretically. They do not show any
sign of the expected odd-even effect (rather the opposite behaviour
is observed  for K!). However, if theoretical ``prejudices'' are
put aside, the situation may not be as bad for Sc and V: indeed, they are
part of the ``low iron group'' elements and their abundances may well follow
the one of Fe, as suggested by current observations. In that case, the 
``odd-even'' effect is overestimated in the theoretical yields adopted here
or those of LSC2000 (Fig. 1). We also noticed  that their solar abundances are 
underproduced by current nucleosynthesis models (Sec. 4.2 and Fig. 6).

\subsection {Fe-peak elements: ~Cr, ~Mn, ~Co, ~Ni, ~Cu and Zn}

The various isotopes of the Fe peak are produced by a variety of 
processes (see WW1995): isotopes with mass number A$<$57 are produced
mainly in explosive O and Si burning and in nuclear statistical equilibrium 
(NSE). Isotopes with A$>$56 are produced in NSE (mostly in
``alpha-rich freeze-out''), but also by neutron captures during hydrostatic
He- or C-burning. Because of the many uncertainties involved in the calculations
(sensitivity to the neutron excess, the mass-cut, the explosion energy etc.)
the resulting yields are more uncertain than for the other intermediate
mass nuclei.

Observations show that the abundance ratio to Fe of Cr, Co, Ni and Zn
is $\sim$solar down to [Fe/H]$\sim$-2.5 to -3. This fact, known already in the
late 80ies, suggests that those elements behave similarly to Fe (at least
in this metallicity range) and, therefore, are produced in a quite similar way.
However, observations in the mid-90ies (Ryan et al. 1996, McWilliam 1997)
show that, as one goes to  even lower metallicities, a
different picture is obtained (see Fig. 7):
Cr/Fe is subsolar and decreasing, while Co/Fe is supersolar and increasing;
the situation is less clear for Ni/Fe, but in all cases the scatter is larger
at very low metallicities  than at higher ones.

For the reasons mentioned in the beginning of Sec. 5, we do not consider
the trends of our models in the range [Fe/H]$<$-3 to be significant.
We do not then attempt here to interpret those recent intriguing findings,
which point, perhaps, to some interesting physics affecting the evolution
of the first stellar generations. We simply notice that such an attempt is made
in Nakamura et al. (1999), who study the sensitivity of the corresponding yields
to various parameters (neutron excess, mass-cut, explosion energy). Their 
conclusion is that the observed Co/Fe excess cannot be explained by any
modification of those parameters.

The yields of WW1995 show a mild metallicity dependence in the case of Cr and
Ni and a more important one in the cases of Mn, Co, Cu and Zn. For that
reason, we obtain different results for those elements between our Cases A and 
B (Fig. 7). The situation for each of those elements is as follows: 

- The Cr/Fe evolution is reproduced satisfactorily for [Fe/H]$>$-2.5; in the disk,
Cr and Fe are produced in similar amounts by SNIa and the Cr/Fe ratio remains 
$\sim$constant.

- Co/Fe decreases steadily as one goes to low metallicities (in Case B). This
trend is not observed in the data and suggests that the ``odd-even'' effect
for that nucleus is overestimated in WW1995; we notice that LSC2000 find
a much smaller effect (Fig. 1).

- The WW1995 yields adequately describe the Ni/Fe evolution, except
at the lowest metallicities ([Fe/H]$<$-3). The LSC2000 yields
would face the same problem, as can be seen in Fig. 1. The excess of Ni/Fe 
obtained in the disk model is due to the overproduction of $^{58}$Ni by the
W7 model of SNIa (see Sec. 4.2).

- The WW1995 yields suggest  a $\sim$constant (solar) Zn/Fe in the halo,
albeit at a value lower than actually observed. On the other hand,
they suggest that Zn/Fe should increase in the disk, while observations
show no such increase. An inspection of the LSC2000 yields in Fig. 1 suggests
that they would face the same problems.

- Finally, the WW1995 yields offer an excellent description of the observed
evolution of Mn/Fe and Cu/Fe. If the observations are correct, we have an
exquisite realisation of the ``odd-even'' effect for Fe-peak nuclei (especially
in the case of Mn), almost a ``text-book'' case. An inspection of the
LSC2000 yields shows that they would do equally well.

\subsection {Fluorine, Neon, Phosphorous, Chlorine, Argon} 

We present in Fig. 7 the evolution of those elements according to
our models, although no observational data exist for them in stars;
fluorine is an exception, its abundance being measured in giants
and barium stars (Jorissen et al. 1992).

We recall that F is produced in WW1995 mainly by neutrino-induced
nucleosynthesis (spallation of $^{20}$Ne)
and the corresponding yields are very uncertain.
As seen in Fig. 1, the F yield of WW1995 are metallicity dependant, 
and this is also reflected in the evolution of the F/Fe ratio 
(Case A vs Case B). We notice again that F may also be produced
in other sites, like in the He-burning shells of AGB stars
(as suggested by the calculations of Forestini and Charbonnel 1997)
or in WR stars. The recent calculations of Meynet and Arnould (2000)
show that the F yields of the latter site are also metallicity 
dependant, but they are important only for metallicities [Fe/H]$>$-1;
at lower metallicities, very few massive stars turn into WR.
Obviously, if AGB and WR stars are the main producers of F, the evolution
of F/Fe ratio may be quite different from the one shown in Fig. 7.

The main Ne isotope is $^{20}$Ne, i.e. Ne should evolve as an $\alpha$-element.
The evolution of Ne/Fe in Fig. 7 is similar to the one of C/Fe. The yields
of WW1995 show a small metallicity dependence (reflected in Case A vs. Case B)
not exhibited by  the yields of LSC2000.

Like Ne, Ar is also an even-Z element. There is no metallicity dependence
in the Ar yields of WW1995 (which explains the similarity between cases A
and B), neither in those of LSC2000. Ar is expected to behave like Si or Ca.

P and Cl are odd-Z elements. When the WW1995 Fe yields are divided by 2,
a $\sim$solar P/Fe and a supersolar Cl/Fe ratio is obtained for halo
stars. In the disk, enhanced P production by massive stars
(due to the ``odd-even'' effect) and by SNIa compensate
for the Fe production by SNIa; as a consequence,
the P/Fe ratio decreases only very slightly.
On the contrary, this compensation does not occur for Cl and the
Cl/Fe ratio decreases in the disk.

\begin{figure*}
\psfig{file=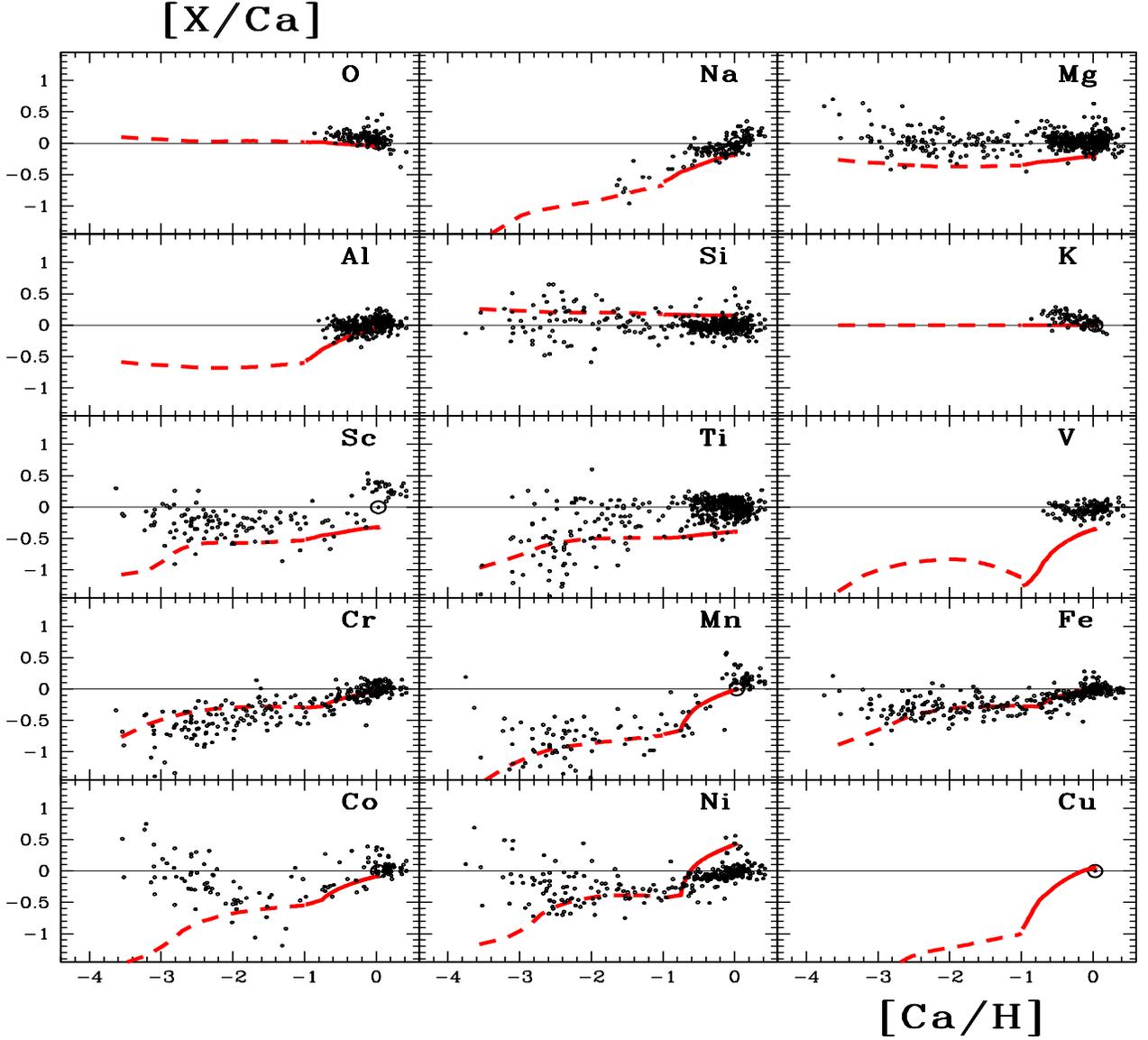,height=16.cm,width=1.05\textwidth}
\caption{\label{} 
Evolution of element/Ca abundance ratios as a function of Ca/H. Observations
are from references listed in Table 1.
Theoretical results ({\it dashed curves} for the halo
and {\it solid curves} for the local disk) are obtained with the
metallicity dependant yields of WW1995 for massive stars and the
W7 and W70 models for SNIa (Iwamoto et al. 1999).
By adopting Ca as a reference element, some of the uncertainties related
to Fe are removed.
 }

\end{figure*}

In the absence of observational data, the nucleosynthesis of 
these elements can not be put on a firm basis. Their solar abundances
are relatively well reproduced with the WW1995 yields (Fig. 6), and
this is quite encouraging. On the other hand, we notice that
the LSC2000 yields show a more pronounced ``odd-even'' effect for P 
and Cl than WW1995.

\subsection {Chemical evolution with respect to Ca}

Traditionally, the results of galactic chemical evolution studies
are presented as a function of Fe/H, i.e. Fe is assumed to play the
role of ``cosmic clock''. However, in view of the uncertainties
on Fe production and evolution (due to mass cut and explosion energy 
in SNII, or to the uncertain evolution of the rate of SNIa), it
has been suggested that Fe should be replaced by a ``robust''
$\alpha$ element, like e.g. O or Ca.

In view of the uncertainties currently affecting the observational status
of oxygen, we choose here Ca as the reference element. Among the data
listed in Table 1 (and plotted in Fig. 7) we selected those including
observations of Ca abundances and we plot the element/Ca ratios in Fig. 8
as a function of Ca/H. We also plot on the same figure the corresponding
model results obtained with the metallicity dependant yields of WW1995
and the W7 model for SNIa (i.e. our Case B).

Several interesting features can be noticed:

- For O, Al, K and V, existing data concern only the disk phase
and are consistent with X/Ca$\sim$solar. Model results  show that
O/Ca and K/Ca ratios are solar over the whole metallicity range; they
also show clearly the ``odd-even'' effect for Al/Ca, V/Fe and Cu/Fe.

- Among the $\alpha$-elements, the observed Mg/Ca and Si/Ca ratios
are solar down to very low metallicities. In our models, we also
find  constant Mg/Ca and Si/Ca ratios, slightly below the observed
values in the former case, and in fair agreement with the observations
in the latter.

- The observed Na/Ca evolution shows clearly the ``odd-even'' effect, especially
with the recent data of Stephens (1999) for the metallicity range 
-1.5$<$[Ca/H]$<$-0.5 and those of
Feltzing and Gustafsson (1998) for [Ca/H]$>$0. This behaviour is fairly well
reproduced by the model.

- The observed Sc/Ca and Ti/Ca ratios are slightly below their solar values
in the halo, with some hint for a decrease of the latter ratio at very low
metallicities. Model results are broadly compatible with those observations.

- Cr/Ca, Fe/Ca and Mn/Ca ratios are all lower than solar in the whole 
metallicity range, exactly as observed. 
The agreement between the model results and the data is
excellent for all three cases, down to the lowest metallicities; notice that
the evolution of Cr w.r.t. Fe was not so well reproduced by the model 
at the lowest metallicities (Fig. 7).

- Finally, the observed Co/Ca and Ni/Ca ratios decrease with decreasing
Ca/H down to [Ca/H]$\sim$-2 and increase at lower metallicities. The former 
trend is rather well reproduced by the model, but not the latter. The
problematic behaviours of Co and Ni at low metallicities do not
disappear when Ca is adopted as ``cosmic clock''.

\section {Alternatives for  Oxygen vs Iron}

In the previous sections we treated oxygen exactly as the other $\alpha$-elements,
i.e. by assuming that[O/Fe]$\sim$0.4$\sim$constant in the halo. However, the
recent intriguing findings of Israelian et al. (1998) and Boesgaard et
al. (1999) suggest that O/Fe continues to rise as one goes from the disk
to halo stars of low metallicities (we shall call these data ``new data''
in this section). Although the observational
status of O/Fe is not settled yet, the ``new data'' certainly call for
alternatives to the ``standard'' scenario to be explored.

\begin{figure*}
\psfig{file=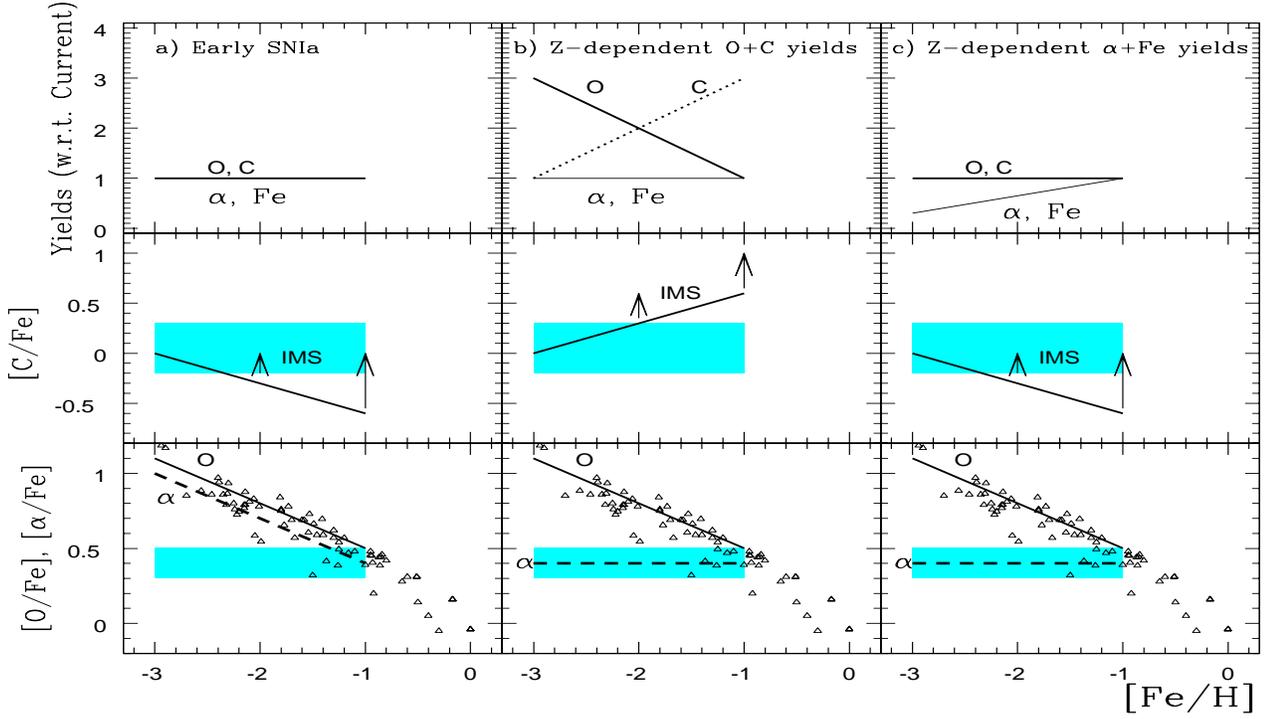,height=10.cm,width=\textwidth,angle=-90}
\caption{\label{} 
Attempts to interpret the ``new data'' of Israelian et al. (1998) and
Boesgaard et al. (1999) on O vs. Fe (appearing in the {\it bottom} panels).
For each scenario discussed in Sec. 6 (presented from left to right),
we show the required modifications in the yields of massive stars 
(w.r.t. their current values, {\it upper panels}), the impact on
the evolution of C/Fe vs. Fe/H ({\it middle } panels) and the
impact on the evolution of O/Fe and $\alpha$/Fe vs Fe/H ({\it bottom}
panels). In the {\it middle} panels, {\it thick solid} lines show the
modified C/Fe evolution, while the {\it shaded area} shows the range
of observed values; {\it arrows} show qualitatively 
the effect of including carbon production 
from intermediate mass (IMS), definitely excluding Case (b).
In the {\it bottom} panels, the {\it thick solid}
line shows the modified O/Fe evolution, the {\it thick dashed} line
the modified $\alpha$/Fe evolution, and the {\it shaded area} represents
schematically current observations of $\alpha$/Fe in the halo.
Scenario (c), on the right, seems to be the only able to 
explain the ``new'' data of Israelian et al. (1998) and
Boesgaard et al. (1999) without violating other obervational constrains.  
For details see Sec. 6.}
\end{figure*}

An obvious alternative is to assume that Fe producing SNIa enter the
galactic scene as early as [Fe/H]$\sim$-3, instead of [Fe/H]$\sim$-1
in the ``standard'' scenario. Indeed, the first white dwarfs, resulting
from the evolution of $\sim$8 M$_{\odot}$ stars, are produced quite
early on in the galactic history; if their companions are almost equally
massive, their red giant winds would push rapidly the white dwarf beyond
the Chandrasekhar mass, and induce a SNIa explosion. The
subsequent evolution of the SNIa rate (not well known today), should then
be such as to ensure a continuous, smooth decline of O/Fe with [Fe/H],
as the ``new data'' suggest. Such a behaviour is indeed obtained in 
the calculations of Chiappini et al. (1999), which have not been adjusted
as to fit the new data: it is a direct consequence of their adopted
formalism for the SNIa rate.

The problem with this ``alternative'' is that it also affects the evolution 
of the other $\alpha$/Fe abundance ratios in the halo. Observationaly,
none of the $\alpha$-elements shows a behaviour comparable to the one
suggested by the ``new data'' for oxygen (see Fig. 7 for Mg, Si and Ca).
The ``new data'' can simply not be explained in terms of SNIa only, because
this would spoil the current nice agreement with the other $\alpha$-elements
(see Fig. 9a).
[{\it Notice}: C/Fe would also decrease with metallicity
quite early  in that case, but this 
is not a serious problem, since C from intermediate mass stars could
keep the C/Fe ratio close to solar, as observed (and indicated in Fig. 9)]. 

A second possibility is that the O yields from massive stars are,
for some reason, metallicity dependant. It is already known that this
happens for the C and N yields of massive stars, for metallicities
Z$>$0.1 Z$_{\odot}$: because of intense stellar winds, the most
massive stars lose their envelope already during He-burning. This
envelope is rich in H-burning products (like
He and N) and later in early He-burning products (essentially C).
Thus, less mass is left in the He-core to be processed into oxygen
(Maeder 1992). As discussed in Sec. 5.1, this metallicity dependence
of C yields from massive (WR) stars, can indeed explain the observed
C/O evolution in the disk. However, Prantzos et al. (1994) have shown that
the effect is clearly negligible for the evolution of oxygen in the disk, 
at least with Maeder's (1992) yields. And at lower metallicities, the
effect is virtually inexistent: even the most massive stars present 
negligible mass losses.
Thus, current models suggest that metallicity dependant Oxygen yields 
cannot help explaining the new data.

However, the effect may have been underestimated. After all, stellar
mass loss is yet poorly understood. Suppose then that, starting
at [Fe/H]$\sim$-3, massive stars
produce less and less oxygen as their metallicity increases, because an
ever larger part of their envelope is removed. Their inner layers,  producing
the other $\alpha$-elements and Fe, are not affected by mass loss; the
resulting $\alpha$/Fe abundance ratio is constant with metallicity, while 
the corresponding O/Fe is decreasing with metallicity. The problem
encountered by the first alternative seems to be solved.
However, in the expelled mass of those stars, the abundances of
He, N and C should be particularly enhanced. The resulting N/Fe and C/Fe
ratios should be steadily increasing with metallicity in the halo
 (see Fig. 9), which is not observed; and introducing N and C from 
IMS would only make things worse. Thus, several arguments suggest that
metallicity dependant oxygen (and, by necessity carbon)
yields of massive stars cannot explain the ``new data''.

A third alternative concerns the possibility of having metallicity
dependant yields of Fe and all elements heavier than oxygen (while
keeping the O,N,C yields independant of metallicity below [Fe/H]$\sim$-1).
In that case, the yields of $\alpha$-elements and Fe would decrease
with decreasing metallicity at the same rate, producing a quasi-constant
$\alpha$/Fe abundance ratio in the halo, as observed. The O/Fe and C/Fe
ratios would both decrease with increasing metallicity (Fig. 9); however, in
the latter case, this decrease would be compensated by C production from
IMS, so that the C/Fe ratio would remain $\sim$constant in the halo,
as observed. Thus, from the three studied alternatives, we think that
only the last one cannot be at present rejected on observational grounds.

What could be the physics behind such a metallicity dependence of the yields
of $\alpha$-elements and Fe in massive stars?
First, we notice that the required effect is very small: a factor of
$\sim$3 increase is required in the yields for a 100-fold increase in
metallicity (between [Fe/H]=-3 and [Fe/H]=-1, see Fig. 9), i.e. of the
same order as the ``odd-even'' effect in Fig. 1. Our scenario requires
that the supernova layers inside the C-exhausted core (i.e. the layers 
containing all the
elements heavier than oxygen) be well mixed during the explosion. Various
instabilities could contribute to that, either in the pre-supernova
stage (in the O-burning shell, Bazan and Arnett 1998) or during the explosion
itself (as in SN1987A, Arnett et al. 1989). This is required in order to
ensure that the $\alpha$/Fe ratio will be $\sim$ constant in the ejecta.
But the main ingredient is that the structure of the star depends on metallicity,
in the sense that lower metallicity cores are more compact than higher
metallicity ones. Then, at the lowest metallicities (say [Fe/H]$\sim$-3), after
the passage of the shock wave, a relatively large proportion of the
well mixed C-exhausted core will fall back to the black hole, feeling a strong
gravitational potential. At higher metallicities,  the core is less
compact and a larger proportion of the C-exhausted core escapes. At all
metallicities, oxygen (and lighter elements as well) 
are located in the losely bound
He-layers and manage always to escape with the same (metallicity independant)
yields.

If the ``new data'' of Israelian et al. (1998) and Boesgaard et al. (1999)
on O vs Fe are confirmed, some radical revision of our ideas on stellar
nucleosynthesis will be required. At present, we think that our third
alternative (schematically illustrated in the right panels of Fig. 9) is
both plausible and compatible with all currently available data.

\section {Evolution of Mg isotopic ratios}

There are very few cases where observations allow to check models
of isotopic abundance evolution in the Galaxy, especially concerning
the early (i.e. halo) phase of that evolution. One of these rare cases
concerns the Mg isotopes $^{25}$Mg and $^{26}$Mg.

All magnesium isotopes are mainly produced by hydrostatic burning
in the carbon and neon shells of massive stars. The production of the
neutron-rich isotopes $^{25}$Mg and $^{26}$Mg is affected by the 
neutron-excess (i.e. their yields increase with initial stellar metallicity)
while $^{24}$Mg is produced as a primary (in principle). Thus, the
isotopic ratios $^{25}$Mg/$^{24}$Mg and $^{26}$Mg/$^{24}$Mg are expected
to increase with metallicity.

\begin{figure}
\psfig{file=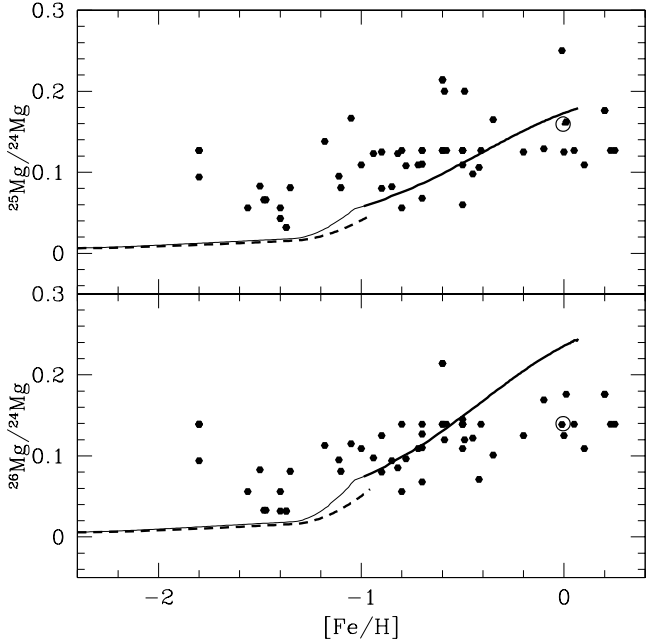,height=9.cm,width=0.5\textwidth}
\caption{\label{} 
Evolution of the isotopic abundance ratios of Mg as a function
of  metallicity [Fe/H]. The {\it upper panel} shows the evolution
of $^{25}$Mg/$^{24}$Mg and the {\it lower panel}  the evolution
of $^{26}$Mg/$^{24}$Mg with respect to [Fe/H]. In both  panels the
{\it solid curve} corresponds to the disk model and the  {\it dashed
curve} to the halo model. 
The observed isotopic ratios 
are  from Gay \& Lambert (1999), McWilliam and Lambert (1988),
Burbuy et al. (1987), Barbuy (1985, 1987), Lambert and McWilliam (1986)
and Tomkin and Lambert (1980). Corresponding solar ratios in both panels
are shown with ${\odot}$.
}
\end{figure}

Observational evidence of a decline of the abundances of $^{25}$Mg and
$^{26}$Mg relative to $^{24}$Mg in low metallicity stars
 was reported as early as 1980 (Tomkin and Lambert 1980). 
In a recent work Gay and Lambert (1999) derived Mg isotopic abundance
ratios for 19 dwarf  stars in the metallicity range -1.8$<$[Fe/H]$<$0,
using high resolution spectra of the MgH A-X 0-0 band at 5140 \AA.
They compared their observations with the theoretical predictions
of Timmes et al. (1995) in the solar neighbourhood and 
found an overall good agreement.

The evolution of Mg isotopic abundance ratios of our models
is plotted as a function of [Fe/H]  in Fig. 10.
The  upper panel represents the evolution of 
$^{25}$Mg/$^{24}$Mg  and the lower panel  the one of $^{26}$Mg/$^{24}$Mg.
Both ratios increase slowly with [Fe/H]. 
$^{25}$Mg/$^{24}$Mg becomes slightly larger than the corresponding
solar ratio at [Fe/H]$\sim$0,
while $^{26}$Mg/$^{24}$Mg is 60\% higher than solar at that metallicity.
This is consistent with the results of Fig. 6 (lower panel), showing that
$^{26}$Mg is produced with its solar value at Sun's formation, while
$^{25}$Mg and $^{24}$Mg are slightly underproduced.  
We notice that Timmes et al. (1995) find also supersolar Mg isotopic 
ratios at [Fe/H]=0, but the $^{26}$Mg excess is not as large as ours.
We think that this difference is due to our use of the Kroupa et al. 
(1993) stellar IMF, favouring the  $^{26}$Mg yields w.r.t those of
$^{24}$Mg; Timmes et al. use the Salpeter IMF.

In Fig. 10 we compare our results with observations from various sources, 
including the recent data of Gay and Lambert (1999).
The observational trends are, globally, reproduced by our model for disk
stars, although
the $^{26}$Mg/$^{24}$Mg ratio is higher than observed for
stars of near  solar metallicity. More interesting is the fact that
the model isotopic ratios are systematically lower than observations
for halo stars (below [Fe/H]$\sim$-1). This was also noticed in Timmes
et al. (1995). It may well be that the WW1995 yields underestimate
the importance of the neutron-excess in the production
of the Mg isotopes at those metallicities.  Another possibility is that
there is some other source of the neutron-rich Mg isotopes in the
late halo, like e.g. AGB stars with He-shells hot enough to
activate the $^{22}$Ne($\alpha$,n)$^{25}$Mg neutron source. This 
reaction, would not only provide neutrons for the s-process in those
stars, but it would also produce large amounts of $^{25}$Mg and 
$^{26}$Mg. At present, the operation of that source in
AGB stars of disk-like metallicities seems improbable, but there 
is no evidence as to what may happen at lower metallicities.

\section{ Summary}

In this work we present a comprehensive study of the evolution of the abundances of
intermediate mass elements (C to Zn) in the Milky Way halo and in the local disk.
We use a consistent model in order to describe the evolution of those two galactic
subsystems. The model assumes strong outflow in the halo phase and slow infall
in the disk, which allow to correctly reproduce the corresponding metallicity
distributions; these observables constitute the strongest constraints for chemical
evolution models of those regions. Also, we consider the halo and the disk to evolve
independently, since there is no hint at present for a physical connection 
between the two (see Sec. 3.3). We note that this type of modelisation has very
rarely been done before.

The second important ingredient of this study is the consistent use of
metallicity dependant yields for all isotopes. We adopt the yields of WW1995 and
we note that there is a remarkably good agreement between them and the more recent ones
of LSC2000 (but also some important differences). Only one study of similar scope
has been done before with the
metallicity dependant WW1995 yields (Timmes et al. 1995),
but it utilised an inconsistent model for the halo. The study of Samland (1998)
used  appropriate models for the halo and the disk, but made several approximations
concerning the stellar lifetimes and the metallicity dependence of the yields.
We note that we have divided the (uncertain, anyway) Fe-peak isotopic 
yields of WW1995
by a factor of 2, in order to obtain abundance ratios w.r.t Fe consistent
with observations; indeed, Timmes et al. (1995) recognised the problem
with the WW1995 Fe yields and presented also results for twice and half the
nominal values.
We also performed calculations with metallicity independant yields (at solar
metallicity only) in order to illustrate the differences with the metallicity
dependant ones. 
In all cases we used the recent yields of Iwamoto et al. (1999)
for SNIa, which are also metallicity dependant (this
dependence affects very little the results).
We only used yields from massive stars and SNIa, in order to find
out for which elements and to what extent is the contribution of other sources
mandatory.

We compared our results to a large body of observational data. In Sec. 4
we ``validated'' our model, by showing that it reproduces in a satisfactory way
all the main observational constraints for the halo and the local disk. We
found that the resulting elemental and isotopic compositions at a galactic age of
9 Gyr compare fairly well to the solar one ; among the few exceptions, the most
important ones concern:

a) The C and N isotopes, which are underproduced. For the major ones ($^{12}$C and
$^{14}$N), both WR and IMS  are candidate sources; for $^{13}$C and $^{15}$N,
IMS and novae are, respectively, the main candidates.

b) The isotopes of Sc, Ti and V, for which there is no other candidate source.
The fact that the corresponding LSC2000 yields are even lower than WW1995
may point to some generic problem of current nucleosynthesis
models for those elements. 

We consider our results for the halo evolution to be significant only above
[Fe/H]$>$-3. The reason is that at lower metallicities 
massive stars have lifetimes comparable to the age of the halo at that point;
since the yields of individual stars are very uncertain, we consider that the
corresponding results have little meaning. Only when the age of the halo becomes
significantly larger than the lifetime of the ``lightest'' massive star
(and ejecta are averaged over the IMF for all massive stars)
we consider our results to become significant. For that reason, we are not 
able to draw any conclusion on the puzzling behaviour of the Fe-peak elements
(Cr, Co, Ni) observed recently below [Fe/H]$\sim$-3.

We have compiled a large number of observational data on the composition of halo
stars. The main conclusions of the comparison of our results to those data 
(Sec. 5 and Figs. 7 and 8) are the following:

- C and N require other sources than those studied here. For C, it could be
WR or low mass stars, contributing to C production in the disk. For N,
the source of primary N required in the halo could be either IMS with hot-bottom
burning or rotationally induced mixing in massive stars.

- The evolution of the $\alpha$-elements
O, Si, S and Ca is well understood (baring the discrepant
``new data'' for O, see below) with the assumption that SNIa contribute most
of Fe in the disk; however, the WW1995 yields underproduce Mg and Ti, and 
inspection of the LSC2000 yields shows that they would not be of help. 

- Similarly, the odd-Z elements Sc and V are underproduced at all metallicities
by both WW1995 and LSC2000 yields; this discrepancy  points to some important
revision required in current models of nucleosynthesis in massive stars,
at least for those elements. It is significant that
observationally, neither Sc nor V show the theoretically expected 
behaviour of odd-Z elements, suggesting that the ``odd-even'' effect may be
overestimated in current nucleosynthesis models.

- Observed abundances of Na and Al also do not show the theoretically
expected behaviour of odd-Z elements, when they 
are plotted w.r.t Fe (Fig. 7).  However, other sources
may be involved in the nucleosynthesis of those two elements
(e.g. H-shell burning in intermediate mass stars in the red giant stage), 
which prevents from drawing  definite conclusions. 
It is remarkable that, when the observed 
Na evolution is plotted vs. Ca (Fig. 8), Na shows indeed the expected 
behaviour of odd-Z element. Observations of Na vs Fe at low metallicities
are necessary to establish the behaviour of this element. 
In the case of Al, NLTE effects play an
important role in estimating its abundance at low metallicities and render
difficult  a meaningful comparison of observations to theory.

- Among the Fe-peak elements, several important discrepancies between theory
and observations are found when results are plotted w.r.t. Fe (Fig. 7).
The theoretical trends of Cr, Co, Ni and Zn deviate from the observed ones to
various extents; in the case of Ni, the adopted W7 model for SNIa largely 
overproduces the main isotope $^{58}$Ni in the disk, as well as$^{54}$Cr,
a minor Cr isotope. We notice that, when results
are plotted w.r.t. Ca (Fig. 8), the observed behaviour of Cr is well 
reproduced by the model; this might imply that it is the Fe yields that
are problematic at low metallicities. We notice that
Cr is produced at layers lying at larger distance from the
core than Fe, and are thus less subject to the uncertainties of the mass-cut.

- There is a remarkably good agreement between the theoretical and the
observed behaviour of the odd-Z Fe-peak elements Mn and Cu, when 
their evolution is plotted  w.r.t. Fe (or w.r.t. Ca, in the case of Mn).

The recent data of Israelian et al. (1998) and Boesgaard et al. (1999)
suggest that oxygen behaves differently than the other $\alpha$-elements.
Although this new picture of O vs Fe is  not confirmed yet, we explored in this
work a few alternatives to the ``standard'' scenario presented here. We thus
showed in Sec. 6 (and Fig. 9), albeit qualitatively only, 
that the only ``reasonable'' way to
accomodate the new data is by assuming that the yields of both
Fe and all $\alpha$-elements (except O, C and He) decrease with decreasing
metallicity for [Fe/H]$<$-1; we also proposed a qualitative explanation for
such a behaviour. 

Finally, we compared the model evolution of the Mg isotopic ratios to current 
observations (Sec. 7 and Fig. 10). 
We found that, although the WW1995 yields of Mg describe 
relatively well the observations in the disk, they systematically
underproduce the halo data. This suggests that the ``odd-even'' effect 
for those isotopes has
been underestimated at low metallicities in WW1995.

In summary, we have revisited the chemical evolution of the halo and the 
local disk with consistent models and metallicity dependant yields of massive
stars and SNIa. We showed that current yields are remarkably successful in 
reproducing a large number of observations, but need revision in
several important cases. For some of those cases, the inclusion
of non-classical ingredients in stellar models (i.e. mass-loss for C, rotationally
induced mixing for primary N) could clearly help, but for most of the others
(Sc, V and Ti at all metallicities, Fe-peak elements at very low metallicities) 
the situation remains unclear. Finally, we explored a few alternatives that 
could help to explain the new O vs Fe data and concluded that viable solutions 
exist, but would require some important modifications of our current
understanding of massive star nucleosynthesis. \par

\medskip
\noindent
{\it Acknowledgements.} Aruna Goswami acknowledges  the hospitality of IAP
(Paris, France)  where  part of the work was being carried out. 
We are grateful to M. Limongi, T. Beers, A. McWilliam, Y. Chen and E. Carretta
for kindly providing
us their data in electronic form.
 This  work is supported by CSIR/CNRS bi-lateral co-operation programme  
No. 19(207)/CNRS/98-ISTAD.

\end{document}